\documentstyle[aps,floats,preprint,epsfig]{revtex}

\begin{document}
\draft
\preprint{\vbox{
\hbox{LMU 14/99}
\hbox{hep-ph/0001074}
\hbox{Jan 2000}}}
\title{Discovery and Identification of 
Extra Gauge Bosons in $e^+e^- \to \nu \bar{\nu} \gamma$}
\author{Stephen Godfrey, Pat Kalyniak and Basim Kamal}
\address{
Ottawa-Carleton Institute for Physics \\
Department of Physics, Carleton University, Ottawa CANADA, K1S 5B6}
 \author{Arnd Leike}
\address{Ludwigs--Maximilians-Universit\"at, Sektion Physik, 
Theresienstr. 37,\\
D-80333 M\"unchen, Germany}

\maketitle

\begin{abstract}
We examine the sensitivity of the process 
$e^+e^- \rightarrow \nu\bar{\nu} \gamma$ 
to extra gauge bosons, $Z'$ and $W'$, 
which arise in various extensions of the standard model.  
The process is found to be sensitive to $W'$ masses up to several TeV, 
depending on the model, the center of mass energy, and the assumed 
integrated luminosity.  If extra gauge bosons were discovered 
first in other experiments,
the process could also be used to measure 
$Z'\nu \bar{\nu}$ and $W'$ couplings. This measurement would provide
information that 
could be used to unravel the underlying theory,
complementary to measurements at the Large Hadron Collider.
\end{abstract}

\section{Introduction}

Extra  gauge bosons, both charged ($W'$) and/or neutral ($Z'$),
arise in many models of physics beyond the Standard 
Model (SM) \cite{c-g,sg}. 
Examples include extended gauge theories such as grand unified 
theories \cite{guts} and Left-Right symmetric models \cite{lrmodels} 
along with the corresponding supersymmetric models, 
and other models such as those with finite size extra dimensions
\cite{led}.
To elucidate what physics lies beyond the Standard Model it is necessary 
to search for manifestations of that new physics
with respect to the predicted particle 
content, both fermions and extra gauge bosons.  Such searches are
a feature of ongoing collider experiments and the 
focus of future experiments.  The discovery of  new particles  would 
provide definitive evidence for physics beyond the Standard Model and, 
in particular, the discovery of new gauge bosons would indicate that 
the standard model gauge group was in need of extension.
There is a considerable literature on $Z'$ searches.  In 
this paper we concentrate on $W'$ searches, for which much less work has 
been done.

Limits have been placed on the existence of new gauge bosons through 
indirect searches based on the
deviations from the SM they would produce in precision electroweak 
measurements. 
For instance, indirect limits from $\mu$-decay constrain the LRM $W'$ 
to $M_{W_{LR}'} \gtrsim 550$ GeV \cite{Barenboim}.  A more severe
constraint 
arises from $K_L - K_S$ mass-splitting which gives $M_{W_{LR}'} \gtrsim 
1.6$~TeV \cite{pdb}. In obtaining
the above limits, it was assumed that the
coupling constants of the two $SU(2)$ gauge groups are equal.

New gauge boson searches at hadron colliders consider their 
direct production via the Drell-Yan process and their subsequent decay 
to lepton pairs. For $W'$ bosons, decays to hadronic jets are 
sometimes also considered.  The 
present bounds on neutral gauge bosons, $Z'$'s, from the CDF and D0 
collaborations at the Tevatron $p\bar{p}$ collider at Fermilab are 
$M_{Z'} > 590 - 690$~GeV with the exact value depending on the 
specific model \cite{pdb}.  For $W'$'s the limits are $M_{W'}> 
300-720$~GeV; again the limits depend on the details of the model 
\cite{pdb}.  The search reach is expected to increase by $\sim 
300$~GeV with 1~fb$^{-1}$ of luminosity \cite{sg,tev-lhc}.  
The Large Hadron Collider is 
expected to be able to discover $Z'$'s up to masses of 
4-5~TeV \cite{sg,tev-lhc}
and $W'$'s up to masses of $\sim 5.9$~TeV \cite{tev-lhc}.  
The $W'$ limits assume SM 
strength couplings and decay into a light stable neutrino 
which is registered in the detector as missing $E_T$.  They can 
be seriously degraded by loosening the assumptions in the model.

In addition, one can place limits on new gauge bosons by 
looking for deviations from SM expectations for observables measured 
at $ep$ and $e^+e^-$ colliders. 

Searches for new gauge bosons at $e^+e^-$ colliders are kinematically 
limited by the available center-of-mass energy so that one searches 
for indirect effects of extra gauge bosons in cross sections
and asymmetries for $\sqrt{s}<M_{V'}$. There is a considerable body of 
work on $Z'$ searches at $e^+ e^-$ colliders and, although the 
discovery limits are very model dependent, they lie in the general 
range of 2-5~TeV for $\sqrt{s}=500$~GeV with 50~fb$^{-1}$
luminosity \cite{sg}. 

In contrast to the $Z'$ case, there are 
virtually no studies of indirect searches for $W'$ bosons at $e^+e^-$
colliders. Recently, 
Hewett suggested that the reaction $e^+e^- \to \nu \bar{\nu} \gamma$ 
would be sensitive to $W'$'s with masses greater than $\sqrt{s}$ 
\cite{hewett}.  In the Standard Model, this process proceeds through 
$s$-channel $Z$ and $t$-channel $W$ exchange with the photon being 
radiated from every possible charged particle.  In extended gauge 
models the process is modified by both $s$-channel $Z'$ and $t$-channel 
$W'$ exchange.  In this paper, we examine this process for various extended 
electroweak models.  The first model we consider is the 
Left-Right symmetric model \cite{lrmodels}
based on the gauge group 
$SU(3)_C \times SU(2)_L \times SU(2)_R \times U(1)_{B-L}$ which has 
right-handed charged currents. 
The second model we consider is the Un-Unified model 
\cite{uum,br} which is based on the gauge group $SU(2)_q \times SU(2)_l 
\times U(1)_Y$ where the quarks and leptons each transform under their own 
$SU(2)$.  The final type of model, which has received considerable
interest 
lately, contains the Kaluza-Klein excitations of the SM gauge bosons which 
are a possible consequence of theories with large extra dimensions
\cite{led}.  
The models under consideration are described in more detail in
Section II.
Additionally, we study discovery limits for various combinations of $W'$
and
$Z'$ bosons with SM 
couplings.  Although these are not realistic models, they have been
adopted as 
benchmarks to compare the discovery reach of different processes.

We will find that, while the process $e^+e^- \to \nu\bar{\nu}\gamma$ can 
indeed extend the discovery reach for $W'$'s significantly beyond 
$\sqrt{s}$, with the exact limit depending on the specific model, it 
is not in general competitive with limits obtainable at the LHC.  
However, if extra gauge bosons are discovered which are not overly 
massive, the process considered here could be used to
measure their couplings. This 
would be crucial for determining the origins of the $Z'$ or $W'$.  As 
such, it would play an important complementary role to the LHC studies.

In the next section we review the relevant details of the various models  
that we use in our 
calculations.  In Section III, we describe the details of our
calculations. The resulting $W'$ discovery limits and projected
sensitivities for 
$W'$ couplings and $Z'\nu\bar{\nu}$ couplings are given in Section IV.  We
conclude with some 
final comments.

\section{Models}

        In this section, we describe the models considered in our
investigation. The so-called Sequential Standard Model (SSM) includes
additional weak gauge bosons of higher mass, with SM couplings. This is a
rather arbitrary scenario which we include only as a benchmark. Since our
emphasis here is on extra $W$'s, we consider a SSM with a $W'$ only, which
we refer to as SSM($W'$), and a
SSM
with both $W'$ and $Z'$, denoted by SSM($W' + Z'$). In the latter, we
will take $M_{Z'}=M_{W'}$
for simplicity.

The general Left-Right symmetric model (LRM) \cite{lrmodels} is based on
the extended
electroweak gauge group $SU(2)_L \times SU(2)_R \times U(1)_{B-L}$.
Left-handed
fermion fields transform as doublets under $SU(2)_L$ and as singlets under
$SU(2)_R$. The reverse is true for right-handed fermions. A right-handed
neutrino is included in the fermion content. The model is parametrized by
the ratio of the coupling constants of the two $SU(2)$ gauge groups, which
we denote as $\kappa = g_R/g_L$. This parameter is allowed to vary here in
the
approximate range $0.55 \lesssim \kappa \lesssim 2.0$
\cite{hewett,pr,cmp}. The lower bound on
$\kappa$
arises from the condition $\sin^{2}\theta_{W} \leq \frac{\kappa^2}{1 +
\kappa^2}$ (or, equivalently,  $\kappa^2 \geq \tan^{2} \theta_{W}$), which
expresses the positivity of a ratio of squared couplings. In principle, 
$\kappa$ is restricted to be less than 1 based on symmetry breaking
scenarios and coupling constant evolution arguments. However, it is
conceivable that this bound may be violated in some Grand Unified
Theory so we take a phenomenological approach and loosen this upper
bound \cite{hewett,cmp}. 

Additionally, a parameter, $\rho$, describes the Higgs content of the
model. If only Higgs doublets are used to break the gauge symmetry to
$U(1)_{em}$, $\rho$ is 1. For Higgs triplets, $\rho$ is 2. A combination
of doublets and triplets leads to an intermediate value of $\rho$ between
1
and 2 \cite{hr}. We will use $\rho = 1$, corresponding to
Higgs doublets.

In the LRM, there is a relationship between the $Z'$ and $W'$ masses, as
follows:

\begin{equation}
\label{massrel}
\frac{M^2_{Z'}}{M^2_{W'}} = \frac{\rho \kappa^2}{\kappa^2 - \tan^2
\theta_W}.
\end{equation}

\noindent The couplings of the extra gauge bosons relevant to our
calculation can be
read from the following
parts of the Lagrangian.

\begin{eqnarray}
\label{LRMLagr}
{\cal L}_{LR} & = & \frac{e \kappa}{\sqrt{2} s_W}
 W'^+_{\mu} \bar{\nu}_R \gamma^{\mu} e_R 
  + \frac{e}{2 s_W c^2_W \sqrt{\kappa^2 - t^2_W}} Z'_{\mu}
\left[ \bar{l} \gamma^{\mu} (1 - \gamma_5) s^2_W (T_{3L} - Q_{em}) l
\right.
\nonumber \\
 & & \left.
+ \bar{l} \gamma^{\mu} (1 + \gamma_5) (\kappa^2 c^2_W T_{3R} -
s^2_W Q_{em}) l \right] + h.c.
\end{eqnarray}
where $e_R = \frac{1}{2}(1+\gamma_5) e$ denotes a right-handed 
electron field.
\noindent Note that we neglect two angles, usually denoted as $\xi$ and
$\zeta$,
which parametrize the $Z-Z'$ and $W-W'$ mixings, respectively. Limits on
these angles are rather severe so this is justified \cite{ls,cln}.
Neglect of these angles implies SM
couplings for the $Z$ and $W$. Additionally, we assume light Dirac-type
neutrinos. 

The Un-Unified model (UUM) \cite{uum,br} employs the alternative
electroweak
gauge
symmetry $SU(2)_q \times SU(2)_l \times U(1)_Y$ with left-handed quarks
and leptons transforming as doublets under their respective $SU(2)$
groups. All the right-handed fields transform as singlets under both
$SU(2)$
groups. The UUM may be parametrized by an angle $\phi$, which represents
the
mixing of the charged gauge bosons of the two $SU(2)$ groups, and by a
ratio $x = (u/v)^2$, where $u$ and $v$ are the vacuum expectation values
of the scalar multiplets which break the symmetry to $U(1)_{em}$.
 The existing constraint on $\phi$ is $0.24 \lesssim
\sin \phi \lesssim 0.99$, based on the validity of perturbation theory.
For $x/\sin^2 \phi \gg 1$, the
$Z'$ mass is approximately equal
to that of the $W'$ and the parameter $x$ may be replaced by $M_{W'}$. The
lepton
couplings of interest to us here arise from the following part of the
Lagrangian.

\begin{eqnarray}
\label{UUMLagr}
{\cal L}_{UU} & = & -\frac{e}{2 s_W} \frac{s_{\phi}}{c_{\phi}}
\left[
\sqrt{2} W'^+_{\mu} \bar{\nu} \gamma^{\mu} l_L + Z'_{\mu} (\bar{\nu}
\gamma^{\mu} \nu_L - \bar{l} \gamma^{\mu} l_L)
\right] + h.c.
\end{eqnarray}

\noindent As expected, the fermion couplings to the additional gauge
bosons are all
left-handed in the UUM. Additional fermions must also be included in order
to cancel anomalies. This is rather difficult to do without generating
flavour changing neutral currents and some considerations of this problem
lead to rather high lower bounds on the $Z'$ mass of about 1.4 TeV
\cite{br}. However, lower $Z'$ and $W'$ masses may be allowed in other
scenarios; hence we take a phenomenological approach in this
investigation.

Finally, we consider the consequences of models which have been of
considerable interest
lately, those containing large extra dimensions \cite{led}. In particular,
we
consider an extension of the SM to 5-dimensions (5DSM) \cite{led2}. The
presence of an
extra dimension of size $R \sim$ TeV$^{-1}$ may imply an infinite tower of
Kaluza-Klein (KK) excitations of the SM gauge bosons. The mass of the
excitations is associated with the compactification scale of the extra
dimension as $n
M_c$ ($n = 1,..., \infty$), where $M_c = 1/R$. The properties of and
relationships among electroweak observables are modified by the presence
of these KK towers. We treat this possibility in a manner similar to the
other models described above; that is, we include in our process the
exchange of a $W'$ and $Z'$ corresponding to the first KK excitations. The
model can be parametrized by an angle $\beta$ which is correlated with the
properties of its Higgs sector, which includes two doublets; for $\sin
\beta \equiv s_{\beta} = 0$, the
SM Higgs may propagate in all 5-dimensions (the bulk) while for $s_{\beta}
= 1$, it is confined to the 4-dimensional boundary. In terms of this
parameter, the physical masses of the lightest electroweak gauge
bosons (corresponding to the experimentally measured masses)
are given, to first order in $M^2_W/M^2_c$, as

\begin{eqnarray}
M^{(ph)2}_W & = & M^2_W \left[1 -  s^4_{\beta} \frac{\pi^2}{3}
\frac{M^2_W}{M^2_c} \right] \\
M^{(ph)2}_Z & = & M^2_Z \left[1 -  s^4_{\beta} \frac{\pi^2}{3}
\frac{M^2_Z}{M^2_c} \right]
\end{eqnarray}

\noindent where $M^2_W = g^2 v^2/2$, as usual.The gauge couplings of the
physical $W$ and $Z$ are also modified by a term of order $M^2_V/M^2_c$.
Global analyses of electroweak parameters put a lower limit on $M_c$ of
about 2.5 TeV so
this is a very small effect. We will therefore neglect it and thus
eliminate
$s_{\beta}$ as a parameter. On the other hand, the fermion coupling of the first
KK excitations, $W'$ and $Z'$, each of mass $M_{V'} = M_c$, is enhanced by
a factor of $\sqrt{2}$. Hence our consideration of the 5DSM amounts to
including a $W'$ and a $Z'$, of equal mass, each coupling as in the SM
apart from an extra factor of $\sqrt{2}$.

\section{Calculation}

The process under consideration is
\begin{equation}
e^-(p_-)+e^+(p_+) \rightarrow \gamma(k) + \nu(q_-) + \bar{\nu}(q_+)\,\, .
\end{equation}
The relevant Feynman diagrams are given in Fig.\ 1. The kinematic 
observables of interest are the photon's energy, $E_\gamma$, and its
angle relative to the incident electron, $\theta_\gamma$, both defined
in the $e^+e^-$ center-of-mass frame. The invariant mass of the
$\nu\bar{\nu}$ pair, $M_{\nu\bar{\nu}}$, and $E_\gamma$ are related via

\begin{equation}
\label{egamma}
E_\gamma = \frac{\sqrt{s}}{2}\left(1-\frac{M_{\nu\bar{\nu}}^2}{s}\right),
\end{equation}
where $s=(p_+ + p_-)^2$.

Let ${\cal M}$ denote the sum of the amplitudes shown in Fig. 1, over
a given number of $Z'$'s and $W'$'s. The doubly differential cross section
is related to $|{\cal M}|^2$ via
\begin{equation}
\label{phasesp}
\frac{d\sigma}{dE_\gamma d\cos\theta_\gamma} = 
\frac{E_\gamma}{2s} \frac{1}{(4\pi)^4} \int_0^\pi
d\theta \sin\theta \int_0^{2\pi} d\varphi |{\cal M}|^2,
\end{equation}
where $\theta$ and $\varphi$ are the polar and azimuthal
angles, respectively, of $q_+$ in a frame where $q_+$ and $q_-$ are
back-to-back. The explicit 
momentum parametrizations are given in the Appendix.

Two approaches to determining $|{\cal M}|^2$ are possible. One can
determine ${\cal M}$ analytically, using spinor techniques 
\cite{CALCUL,BerGiele}
for instance, then square it numerically or one can find $|{\cal M}|^2$
analytically. We have followed both approaches, which provides an
independent
check. Obtaining $|{\cal M}|^2$ analytically has been done both via the
trace method, using the symbolic manipulation program FORM \cite{FORM}, 
and by squaring the helicity amplitudes and summing over
the final state helicities. The latter approach leads to a rather compact
result which we present below.

In order to present $|{\cal M}|^2$, we define the following kinematic
variables. We follow the notation of \cite{BerBurg}, where the SM
contribution for this process was calculated:
\begin{equation}
\begin{array}{rrlrl}
\nonumber
& s =& (p_+ + p_-)^2, & s' =& (q_+ + q_-)^2, \\
\nonumber  & t =& (p_+-q_+)^2, & t' = & (p_- - q_-)^2, \\
\nonumber &  u = & (p_+-q_-)^2, & u' = & (p_- - q_+)^2, \\
\nonumber & k_\pm = & 2 p_\pm \cdot k, & k'_\pm = & 2 q_\pm \cdot k, \\
Z_i = s' - M_{Z_i^2} + i M_{Z_i} \Gamma_{Z_i}, & 
W_i = & t-M_{W_i^2}, & W_i' = & t'-M_{W_i^2}.
\end{array}
\end{equation}
The decay width of the extra neutral gauge boson, $\Gamma_{Z_i}$, into
fermion- antifermion pairs is calculated
in each of the models we consider. We include the one-loop QED, three-loop
QCD and $O(M_t^2/M_{Z'}^2)$ corrections, although their effect on the
cross section
is negligible.
In the following, we denote generalized couplings as may be inferred from
the vertices
\begin{eqnarray}
\label{Zcoup}
Z_i f\bar{f} &=& \frac{ig}{2c_W} \gamma^\mu \left( \frac{1-\gamma_5}{2}
\,\,a_{Z_i}^f + \frac{1+\gamma_5}{2}\,\, b_{Z_i}^f \right) \\
\label{Wcoup}
W_i l\nu &=& \frac{ig}{\sqrt{2}} \gamma^\mu \left( \frac{1-\gamma_5}{2}
\,\,a_{W_i} + \frac{1+\gamma_5}{2} \,\,b_{W_i} \right) .
\end{eqnarray}
Thus, in the SM, $a^e_{Z_1} = 2s_W^2-1$, $b^e_{Z_1}=2s_W^2$, $a^\nu_{Z_1}
= 1$, $b^\nu_{Z_1} = 0$, $a_{W_1}=1$, and
$b_{W_1}=0$.

It is only necessary to present the unpolarized squared amplitude as the
individual polarized contributions may be inferred from the coupling
structure. The spin-averaged unpolarized $|{\cal M}|^2$ is given by:
\begin{eqnarray}
\label{m2}
\nonumber
|{\cal M}|^2_{\rm unp} &=& \frac{(4\pi)^3\alpha^3}{8 s_W^4 k_+k_-}
 \left\{ \mbox{\raisebox{4ex}{}} \right. 
\frac{3s'}{c_W^4} \sum_{\stackrel{\scriptstyle i=1,nz}{j=i,nz}} Z_{ij} 
  [(a^e_{Z_i}a^e_{Z_j}a^\nu_{Z_i}a^\nu_{Z_j} 
+ b^e_{Z_i}b^e_{Z_j}b^\nu_{Z_i}b^\nu_{Z_j})(u^2+u'^2)  \\
\nonumber && \,\,\,\,\,\,\,\,\,\,\,\,\,\,\,\,\,\,\,\, 
\,\,\,\,\,\,\,\,\,\,\,\,\,\,\,\,\,\,\,\, 
\,\,\,\,\,\,\,\,\,\,\,\,\,\,\,\,\,\,\,\, 
\,\,\,\,\,\,\,
+ (a^e_{Z_i}a^e_{Z_j}b^\nu_{Z_i}b^\nu_{Z_j} 
+ b^e_{Z_i}b^e_{Z_j}a^\nu_{Z_i}a^\nu_{Z_j})(t^2+t'^2)] \\
\nonumber &&
+ \frac{4}{s'} \sum_{\stackrel{\scriptstyle i=1,nw}{j=i,nw}} W_{ij}
[(a^2_{W_i}a^2_{W_j}+b^2_{W_i}b^2_{W_j})
(u^2+u'^2) + 2 a_{W_i}a_{W_j}b_{W_i}b_{W_j}(s^2+s'^2)] \\
\nonumber &&
+ \frac{4}{c^2_W} \sum_{\stackrel{\scriptstyle i=1,nw}{j=1,nz}}
[\left( WZ \right)_{ij} (u^2 a^2_{W_i} a^e_{Z_j} a^\nu_{Z_j}
+ u'^2 b^2_{W_i} b^e_{Z_j} b^\nu_{Z_j})  \\
 && 
\,\,\,\,\,\,\,\,\,\,\,\,\,\,\,\,\,\,\,\, 
\,\,\,\,\,\,\, 
+ \left( WZ \right)'_{ij} (u'^2 a^2_{W_i} a^e_{Z_j} a^\nu_{Z_j}
+ u^2 b^2_{W_i} b^e_{Z_j} b^\nu_{Z_j})] 
 \left. \mbox{\raisebox{4ex}{}} \right\},  
\end{eqnarray}
where
\begin{equation}
\begin{array}{rclrcl}
Z_{ij} &=& {\displaystyle {\rm Re} 
\left(\frac{2-\delta_{ij}}{Z_iZ^*_j}\right)}, &
W_{ij} &=& (2-\delta_{ij}) {\rm Re} (F_{W_i} F^*_{W_j}), 
\,\,\,\,\,\, \left( WZ \right)_{ij} = 
{\displaystyle {\rm Re} 
\left(\frac{F_{W_i}}{Z_j}\right)},  \\
&&&&& \\
\left( WZ \right)'_{ij} &=& 
{\displaystyle {\rm Re} 
\left(\frac{F_{W_i}}{Z^*_j}\right)},  &
F_{W_i} &=& {\displaystyle \frac{s'}{W'_i} - \frac{s'k_+ - tk'_- +
uk'_+ 
- 4i\varepsilon(q_+q_-p_+k)}{2 W_i W'_i}}, 
\end{array}
\end{equation}
using the notation $\varepsilon(p_1p_2p_3p_4) = 
\varepsilon_{\mu\nu\rho\sigma} p_1^\mu p_2^\nu p_3^\rho p_4^\sigma$,
where $\varepsilon_{\mu\nu\rho\sigma}$ is the completely antisymmetric
Levi-Civita tensor and $\varepsilon_{0123}=1$.
In Eq.~(\ref{m2}) we have assumed lepton universality with regards to
the $Z'\nu\bar{\nu}$ couplings.
Although it may not be immediately apparent, the contribution to the 
cross section from the state where the $e^-$ and $e^+$ are both 
left-handed is equal to the contribution from the state where they are
both right-handed and the sum is given by the term in Eq.~(\ref{m2}) 
proportional to $a_{W_i}a_{W_j}b_{W_i}b_{W_j}$.  

A relation which was quite useful in simplifying $|{\cal M}|^2$ is
\begin{equation}
\frac{s'}{W'_i} - \frac{s'k_+ - tk'_- + uk'_+ 
- 4i\varepsilon(q_+q_-p_+k)}{2 W_i W'_i} = 
\frac{s'}{W_i} - \frac{s'k_- - t'k'_+ + u'k'_- 
- 4i\varepsilon(q_-q_+p_-k)}{2 W_i W'_i}.
\end{equation}
In the SM limit, Eq.~(\ref{m2}) agrees with the expression given in
\cite{BerBurg} after correcting for the known missing factors of
$1/s'$ in \cite{BerBurg} required on dimensional grounds.

The calculation of $d\sigma/dE_\gamma d\cos\theta_\gamma$ may be performed
analytically or numerically. We have followed both approaches and verified
numerical agreement. Further checks were performed using the program
CompHEP \cite{CompHEP}.

\section{Results}

Before discussing the discovery limits obtained in the various models,
we  present the total cross sections and the differential cross
sections $d\sigma/dE_\gamma$ and $d\sigma/d\cos\theta_\gamma$. In doing so,
all the essential features are illustrated. We take the SM inputs
$M_W = 80.33$ GeV, $M_Z = 91.187$ GeV, 
$\sin^2\theta_W = 0.23124$, $\alpha=1/128$, $\Gamma_Z=2.49$ GeV
\cite{pdb}.
 Since we work only to leading order in $|{\cal M}|^2$,
there is some arbitrariness in what to use for the above input,
in particular $\sin^2\theta_W$.

Kinematically, the maximum allowed value for $E_\gamma$ is $\sqrt{s}/2$.
In addition, to take into account detector acceptance,
$E_\gamma$ and $\theta_\gamma$ have been restricted to the ranges
\begin{equation}
\label{accept}
E_\gamma \geq 10 \,\,{\rm GeV}, \hspace{1cm} 
10^0 \leq \theta_\gamma \leq 170^0.
\end{equation}
The cuts also serve to remove 
the singularities which arise when the emitted photon is soft or collinear
with the beam. Further, we restrict the photon's transverse momentum
to 
\begin{equation}
\label{ptcut}
p_T^\gamma > \frac{\sqrt{s} \sin\theta_\gamma \sin\theta_v }
{\sin\theta_\gamma + \sin\theta_v},
\end{equation}
where $\theta_v$ is the minimum angle down to which the veto detectors
may observe electrons or positrons. We take $\theta_v = 25$ mrad. This
cut has the effect of removing the largest background to our process,
namely radiative Bhabha-scattering where the scattered $e^+$ and $e^-$
go undetected down the beam pipe.

This study was performed in leading order, but
QED corrections to $e^+e^-\rightarrow \nu\bar\nu\gamma$ must be taken
into account in a precision analysis of real data.
They have been known  to $O(\alpha)$ for some time \cite{BerBurg}.
See \cite{qedrev} for a short review {\it of} and further references
{\it to} higher order QED corrections, and \cite{nicrosini} for a
description of a related MC generator. Since our aim is to determine
the statistical power of the process in discovering $W'$'s, there is
no need to include in this study the radiative corrections which will
only marginally influence the number of events.  Complete consistency
at NLO, however, would require determination of the bremsstrahlung corrections
to the {\em generalized} expression (\ref{m2}) and corresponding loop
graphs.

As well, we do not explicitly take into account any higher order
backgrounds. 
A background, which cannot be suppressed, comes from the reaction 
$e^+e^-\rightarrow \nu\bar\nu\nu'\bar\nu'\gamma$. The authors of
\cite{dittmaier} have provided the following cross sections of relevance
here:
$\sigma(e^+e^-\rightarrow\nu_e\bar\nu_e\nu_e\bar\nu_e\gamma)\equiv\sigma_{eeee}
=6.65(2)\,{\rm fb},\ \ 
\sigma_{ee\mu\mu}=7.79(2)\,{\rm fb},\ \ 
\sigma_{\mu\mu\mu\mu}=0.690(2)\,{\rm fb}\ \ $ and\ \ 
$\sigma_{\mu\mu\tau\tau}=1.383(3)\,{\rm fb}$. These results are
for the same conditions as in Table~1 of \cite{dittmaier} but for
$\sqrt{s}=500\,{\rm GeV}$.
The cuts used in obtaining the above numbers differ from ours.
Nonetheless, these cross sections give an idea of the magnitude of the
background.
Assuming lepton universality, the total cross section is $25\,{\rm fb}$
for the process $e^+e^-\rightarrow \nu\bar\nu\nu'\bar\nu'\gamma$.
Imposing our $p_T^\gamma$ cut will suppress it even further.
This background must be included in an 
``$e^+e^-\rightarrow \gamma$ + nothing'' analysis of real data.
We expect that the cross sections of 
$e^+e^-\rightarrow \nu\bar\nu\nu'\bar\nu'\gamma\gamma$ 
and of 
$e^+e^-\rightarrow\nu\bar\nu\nu'\bar\nu'\nu''\bar\nu''\gamma$ 
are so small that they need not be taken into account in the analysis. 

The errors generated from the subtraction of the above backgrounds form
part of the systematic error. As the backgrounds themselves are much smaller
than the signal, though comparable to the new physics effect,
we expect that the error in the SM prediction of the 
backgrounds would be much smaller than the systematic errors arising from
detector and beam uncertainties. We shall return to the issue of systematics
in connection with their influence on the discovery limits presented in 
the next section.

We have calculated three distinct total cross
sections: unpolarized: $\sigma$, for left-handed $e^-$: $\sigma_L$, and
for
right-handed $e^-$:
$\sigma_R$. Fig.\ 2 shows all three plotted versus
$\sqrt{s}$, with $\sigma_L$ and $\sigma_R$
calculated using 100\% beam polarization.
Results are shown for the SM, LRM ($\rho=\kappa=1$), UUM ($\sin\phi=0.6$),
SSM($W'$), SSM($W'+Z'$) and KK model, with $M_{W'}=750$ GeV in each case.
These mass and coupling parameter choices are rather arbitrary, made to
illustrate general behaviour.
It is worth noting at this point that in the UUM and SSM($W'+Z'$),
the correction to the SM cross section changes sign as $\sqrt{s}$ is
varied. This arises, for certain $\sqrt{s}$ and $M_{W'}$, due to a
negative
interference term between the SM and $Z'/W'$ diagrams in these models.

It is clear from the presence of the peaks in Fig.\ 2 that we are
also probing $Z'$'s, in those models which include them. (There is
also a very sharp peak at lower $\sqrt{s}$,
off the plot, due to the SM $Z$.) The $Z'$
peaks generally occur for $\sqrt{s}$
slightly above the $Z'$ mass since the photon carries away some of the
energy. At very high energies, the SM $Z$
contribution is negligible. Further, by using a right-handed $e^-$ beam,
we can reduce
the SM $W$ contribution (depending on the degree of polarization).
Then we directly probe the $W'$ (and $Z'$) in the LRM, while in the 
SSM($W'+Z'$) and KK model, we probe only the $Z'$. The latter two models
as well as the two remaining models all require some component of 
left-handed polarization to probe the $W'$. The above features are borne
out in Fig.\ 2.

In order to see which  regions of $E_\gamma$ are most sensitive to the new
physics, we plot for left- and right-handed electron beams respectively,
in Figs. 3(a) and 4(a) $d\sigma/dE_\gamma$ versus $E_\gamma$
and in Figs. 3(b) and 4(b)
the deviation from the SM result divided by
the square root of the predicted cross section versus $E_\gamma$. 
We show results for $\sqrt{s} = 500$ GeV with 100\%  $e^-$ beam
polarization in these figures.

First, we note the shape of $d\sigma/dE_\gamma$ in Figs 3(a) and 4(a).
For left-handed electrons, the bulk of the cross section comes from
the low $E_\gamma$ region; the reduction at very low $E_\gamma$ is
due to the $p_T^\gamma$ cut and the sharp peak at $E_\gamma \simeq 240$
GeV is due to the radiative return to the $Z$ pole. For 100\% right
polarized electrons,
the cross section is rather flat in the low to moderate $E_\gamma$ region,
then increases as a result of 
the $Z$ peak at high $E_\gamma$. On the other hand, since the 
right-handed cross section is two orders of magnitude smaller than
the left-handed cross section away from the $Z$ peak, any realistic degree
of polarization (i.e.\ 90\%) will lead to a large contribution from
$\sigma_L$ to the
low $E_\gamma$ region. In general, there can also be a peak due to a
$Z'$ for $M_{Z'}<\sqrt{s}$ which occurs at
\begin{equation}
\label{peak}
E_\gamma^{\rm peak} = \frac{\sqrt{s}}{2}\left(1-\frac{M_{Z'}^2}{s}\right)
\end{equation}
in analogy with the SM $Z$.

Most important, however, is the relative statistical significance, shown
in Figs. 3(b) and 4(b). In both the left- and right-handed cases, the 
low $E_\gamma$ region is the most sensitive to the new physics. There are
two reasons for this. First, for left-handed electrons, the cross section
is largest at low $E_\gamma$, as mentioned above. Second, the lower
$E_\gamma$, the higher the mass probed in the $Z'$ propagator via 
Eq.\ (\ref{egamma}).
The relative effect is even larger when
combining the $\chi^2$'s from the different bins, since it is the squares
of the plotted quantities which will enter.
Overall, the KK model leads to the most statistically significant deviations,
except for the 100\% left polarized case where the SSM($W'$) exhibits the
largest deviation. We can also see clearly how the sign of the deviation
from the SM
depends on the beam polarization. For the KK model and SSM($W'+Z'$),
we observe a negative deviation with right-handed polarization, 
implying a negative $Z'$ contribution, versus a positive 
overall contribution coming from the left-handed channel.
Clearly, interference effects will make probing $W'$'s nontrivial.
We shall return to this point in the next section.

In Figs. 5 and 6 we plot the analogous quantities relevant to
$d\sigma/d\cos\theta_\gamma$, versus $\cos\theta_\gamma$. We note that
both $d\sigma/d\cos\theta_\gamma$ and the relative statistical significance
are peaked in the forward and backward directions and both are very
nearly symmetric in $\cos\theta_\gamma$. The latter implies that the  
forward-backward asymmetry will be small and, therefore, the
deviation from the SM 
forward-backward asymmetry will also be small, at least in absolute
magnitude. We therefore do
not expect the forward-backward asymmetry to serve as a useful probe
of the new physics, which is confirmed by explicit calculation. An
important observation is that our $p_T^\gamma$
cut, while eliminating a large background, has also eliminated much
of our signal (both from the small angle and soft events) which
was appreciably stronger prior to the cut. A more detailed study,
including a detector simulation, would
be required to determine whether the background could be accurately
subtracted with a looser $p_T^\gamma$ cut.

\subsection{Discovery Limits for $W'$'s}

The best discovery limits were in general obtained using the observable
$d\sigma/dE_\gamma$, combined with beam polarization, while 
$d\sigma/d\cos\theta_\gamma$ was less sensitive.
Comparable or equal limits were obtained using the total cross section,
with an additional cut on the energy to eliminate
the $Z$ pole radiative return events:
\begin{equation}
\label{emax}
E_\gamma^{\rm max} = \frac{\sqrt{s}}{2}\left(1-\frac{M^2_{Z}}{s}
\right)-6\Gamma_{Z}.
\end{equation}
As can be seen from Figs.\ 3(b) and 4(b), the $Z$ pole region is quite 
insensitive to new physics. In the cases that $d\sigma/dE_\gamma$ provided
a better limit than the total cross section, the improvement was of order
50 GeV. However, the $\chi^2$
obtained using the total cross section is a somewhat less stable
function
of $M_{W'}$ as the sign of the deviation from the SM cross
section may change with $M_{W'}$
leading to isolated regions of insensitivity at low
$M_{W'}$. Also, when systematic errors are included, the limits obtained
using $d\sigma/dE_\gamma$ are affected much less than those obtained
using the total cross section.

Substantially weaker limits were obtained using the left-right 
asymmetry,
\begin{equation}
\label{alr}
A_{LR} = \frac{\sigma_L-\sigma_R}{\sigma_L+\sigma_R},
\end{equation}
even when including systematic errors only one 
half those used in the $d\sigma/dE_\gamma$ calculation (since one expects 
some cancellation of errors between the numerator and denominator in 
$A_{LR}$). 
As expected from the discussion of the previous section, the 
forward-backward asymmetry, $A_{FB}$, was quite insensitive to the new 
physics. In light of the above, we restrict the remaining discussion 
to limits obtained using $d\sigma/dE_\gamma$ as an observable.

In obtaining the $\chi^2$ for $d\sigma/dE_\gamma$, we used 10 equal
sized energy bins in the range $E_\gamma^{\rm min} < E_\gamma <
E_\gamma^{\rm max}$, where $E_\gamma^{\rm min}$ follows from the 
$p_T^\gamma$ cut Eq.~(\ref{ptcut}):
\begin{equation}
E_\gamma^{\rm min} = \frac{\sqrt{s} \sin\theta_v}{1 + \sin\theta_v},
\end{equation}
which supersedes the acceptance cut of Eq.~(\ref{accept}). We have
\begin{equation}
\chi^2 = \sum_{\rm bins} \left(\frac{d\sigma/dE_\gamma - d\sigma/dE_{\gamma, 
{\rm SM}}}{\delta d\sigma/dE_\gamma}\right)^2,
\end{equation}
where $\delta d\sigma/dE_\gamma$ is the error on the measurement and
analogous formulae hold for other observables. One sided 95\% confidence
level discovery limits are obtained by requiring $\chi^2\geq 2.69$
for discovery. Systematic errors, when included,
were added in quadrature with the statistical errors.

In determining the limits for the case of polarized electron beams, we
show results for the polarization state which in general
has the largest sensitivity (deviation from the SM) for a given model; a
right-handed $e^-$ beam for the LRM and a left-handed
beam for all other models.
We used one half the unpolarized luminosity
for the polarized case, assuming equal running time in each
polarization state.

The discovery limits for all five models are listed in Table I,
for $\sqrt{s}=0.5$, 1.0 and 1.5 TeV, using the same input parameters
as for the cross sections presented in the previous section. We show
limits for both an unpolarized $e^-$ beam and for a 90\% polarized one.
For each center-of-mass energy, two luminosity scenarios are considered
and we present
limits obtained with and without systematic errors. Our prescription is
to include a 2\% systematic error per bin. This number is quite
arbitrary but seems reasonable, if not conservative, considering the 
clean final state. In addition to detector systematics, which we
expect will dominate, there are uncertainties associated with the
beam luminosity and energy, which will be spread over a range. The 
systematic errors associated with the background subtraction should be 
much smaller than 2\% as should be the errors in the calculation of the
QED corrections. The 2\% number should not be taken too seriously 
therefore, except to highlight the fact that a precision measurement is
required to take full advantage of the large event rate.

Certain features are common to all models. With no systematic error 
included, we observe quite an improvement in the limits with increased
luminosity. The only exception is the UUM at $\sqrt{s}$ of 1.5 TeV,
where the improvement is minimal. The reason is that the $\chi^2$ 
decreases very rapidly as $M_{W'}$ is increased in the vicinity of
the limit, hence increasing the luminosity by a factor of 2.5 does little.
The unusual $\sqrt{s}$ dependence can be attributed to the 
interference effect noted in the previous section,
which results in, for example, for the UUM 
with $\sin\phi=0.6$ and an integrated luminosity of 500 fb$^{-1}$, 
a lower discovery limit at $\sqrt{s}=1.5$ TeV than at 0.5 and 1 TeV.
We will return to this peculiar behaviour later in the section.
When 2\%
systematic errors are included, the high luminosity scenario yields
little improvement in the limits in any of the models, since the
systematic error now dominates
the statistical.

Perhaps surprising at first is the observation that 90\% beam polarization
does not improve the limits very much. This follows from taking
into account the reduced luminosity and the fact that the left-handed
component tends to dominate the unpolarized cross section by a
considerable amount. On the other
hand, we observed that if the polarization is pushed beyond 90\%, then
the right-polarized limits can increase significantly in those models
in which the beyond-SM bosons have a non-zero right-handed coupling: the
LRM, KK model and SSM($W'+Z'$).
In the latter two models, it is however, the $Z'$ which is being probed.
The higher degree of polarization is required to eliminate the
contamination from the much larger left-handed component. Thus, the
primary
advantage of beam polarization is to distinguish between models
and measure the new couplings, as will be investigated in the next
section.

Fig.\ 7 presents the $W'$ mass
discovery limits obtainable in the LRM with an unpolarized beam,
plotted versus $\kappa$ for $\rho=1$ and $\sqrt{s}=$ $0.5$, 1.0, 1.5
and 2 TeV using a luminosity of 50 fb$^{-1}$ for $\sqrt{s}=0.5$ TeV and 
200  fb$^{-1}$ for the higher energies.
Only statistical errors are included.
Depending on 
$\sqrt{s}$ and  $\kappa$, the limits range from 0.8 to 2.8 TeV.
We expect greater deviations from the SM, and hence larger limits, as
$\kappa$ is increased since this increases the $W'$ coupling strength,
as can be seen from Eq.~(\ref{LRMLagr}).
The predicted dependence on $\kappa$  is generally observed,
except at low $\kappa$ where we notice a moderate {\em increase} in the
limits, even though the $W'$ couplings have weakened. We attribute this
effect to the $Z'$, whose couplings are enhanced (but its mass increased)
in the low $\kappa$ region. This was indicated by an appreciable 
improvement in the limits for low $\kappa$ and $\rho=1$
versus those obtained using $\rho=2$ and consequently a heavier 
$Z'$, via Eq.~(\ref{massrel}). 
Fig.\ 8 demonstrates
the improvement in bounds in the moderate to large $\kappa$ region
obtained when a 90\% or 100\%
polarized right-handed $e^-$ beam is used. The
beam polarization picks out the LRM $W'$ and suppresses the SM $W$.
Fig.\ 8(a) shows that for $\kappa>1$, 90\% beam polarization improves
the limits. Further increasing the polarization leads to substantial
improvements, even at lower $\kappa$, as demonstrated in Fig.\ 8(b).

The dependence of the limits in the UUM on $\sin\phi$ is shown in
Fig.\ 9, for $\sqrt{s}=$ $0.5$, 1.0, 1.5 and 2.0 TeV,
under the same running conditions as Fig.\ 7.
Only the unpolarized case is considered as beam polarization was not
beneficial. Again, only statistical errors are included.
At each $\sqrt{s}$, we note that the contour defining the exclusion 
region as a function of $\sin\phi$ is a complicated curve.
The consequence is that for $\sqrt{s}=1$ TeV,
we obtain better limits over a range of $\sin\phi$ than we do for
$\sqrt{s}=1.5$ and even  $\sqrt{s}=2$ TeV. 
Essentially, this is due to the complicated interference with the SM
diagrams.
In general, as $\sin\phi$ increases, the UUM couplings also
increase, as can be seen from Eq.~(\ref{UUMLagr}), so that higher
mass scales are probed. So, referring to Fig. 2, 
the peak in the cross section (due to
the $Z'$) at the scale being probed shifts to the right. 
But 
the sign of the deviation from the SM changes with  $\sqrt{s}$ for 
fixed $M_{W'}$ (or vice-versa) such that the UUM cross section
dips below the SM over some region to the left of the peak, then goes
back above it for small $\sqrt{s}$ (or large $M_{W'}=M_{Z'}$ for fixed
$\sqrt{s}$). Hence, there is a small step in the limits near
$M_{W'}=\sqrt{s}$, corresponding to passing the rightmost crossing with the
SM and another structure in the contour
at some higher $M_{W'}$ such that the leftmost crossing
is situated near $\sqrt{s}$. One sees this explicitly by plotting $\chi^2$
versus $M_{W'}$ for fixed $\sqrt{s}$ and $\sin\phi$ and observing a 
dip in the $\chi^2$ at relatively low $M_{W'}$. Had we used $\sigma$ as
an observable, the dip would be much more pronounced since 
$\sigma-\sigma_{\rm SM}$ passes through zero, but 
$d\sigma/dE_\gamma - d\sigma/dE_{\gamma, {\rm SM}}$ may differ in sign
between bins, leading to a nonzero $\chi^2$ at the crossing points.
Once $\sin\phi$ is large enough that we
are probing the region
to the left of the leftmost crossing, the limits shoot up in 
an impressive fashion as the dip in  $\chi^2$ never goes back down
to 2.69.
The shape of the plot is
luminosity dependent since, as pointed out earlier in this section, the degree
to which increased luminosity improves the limits depends on the rate
at which the $\chi^2$ decreases with increasing $M_{W'}$ in the 
vicinity of the limit. That,
in turn, varies with $\sqrt{s}$ for fixed $\sin\phi$ and with
$\sin\phi$ for fixed $\sqrt{s}$.

\subsection{Constraints on Couplings}

In this section, we consider constraints
which can be put on the couplings of extra gauge bosons by the process 
$e^+e^-\rightarrow\nu\bar\nu\gamma$.
These constraints are significant only in the case where 
the mass of the corresponding extra gauge boson is 
considerably lower than its search limit in this process.
In most models, the process $e^+e^-\rightarrow f\bar f$ and/or searches
at the LHC are
more sensitive to a $Z'$ or $W'$ (LHC) than the process 
$e^+e^-\rightarrow\nu\bar\nu\gamma$.
We assume here that a signal for an extra gauge boson has been detected 
by another experiment.

Given such a signal, we derive constraints (at 95\% C.L.) on the
couplings of 
extra gauge bosons. We present the constraints in terms of couplings
normalized as follows relative to Eqs.~(\ref{Zcoup}) and (\ref{Wcoup}).
\begin{equation}
\begin{array}{rlrl}
L_f(Z)  = & \frac{g}{4c_W}a^f_{Z_i} & \;\;\; R_f(Z)  = &
\frac{g}{4c_W}b^f_{Z_i} \\
L_f(W)  = & \frac{g}{2\sqrt{2}}a_{W_i} & \;\;\; R_f(W)  = &
\frac{g}{2\sqrt{2}}b_{W_i}.
\end{array}
\end{equation}

The constraints correspond to 
\begin{equation}
\label{chi2}
\chi^2=\sum_i\left(\frac{O_i(SM)-O_i(SM+Z'+W')}{\delta O_i}\right)^2=5.99,
\end{equation}
where $O_i(SM)$ is the prediction for the observable 
$O_i$ in
the SM, $O_i(SM+Z'+W')$ is the prediction of the
extension of 
the SM and $\delta O_i$ is the expected experimental error.
The index $i$ corresponds to different observables such as
$\sigma$ and
$A_{LR}$.

Our assumptions concerning beam polarization are as follows. For single
beam ($e^-$) polarization, we assume, as in the previous section,
equal running in left and right polarization states. For double beam
polarization, we assume equal running in the $LR$ and $RL$ states, but 
no running in the $LL$ and $RR$ states. Thus,
\begin{equation}
A_{LR} = \frac{\sigma_{LR}-\sigma_{RL}}{\sigma_{LR}+\sigma_{RL}},
\,\,\,\,\,\, \mbox{$e^-$ and $e^+$ polarized},
\end{equation}
where the first subscript of $\sigma$ refers to the $e^-$ helicity.
Note that for 100\% polarized $e^-$ and $e^+$, 
$\sigma_{LL}=\sigma_{RR}=0$ in all the models we consider. This remains
approximately valid as the couplings deviate from their model-defined
values.

In Figs.~\ref{nnng1} and \ref{nnng2}, 
we present $Z'\nu\bar{\nu}$ coupling constraints
assuming there is no signal for a $W'$. This is the
case when the SM is extended by $U(1)$ factors only.
It can also happen in models where the $W'$ has purely right-handed
couplings and the right-handed neutrino is heavy.
Then, the process $e^+e^-\rightarrow\nu\bar\nu\gamma$ would be
one of the best
for constraining the couplings of the $Z'$ to SM neutrinos {\it below} the 
$Z'$ resonance.
If there is also a signal for a $W'$, a similar analysis could be 
performed including
the $W'$ parameters, as measured in other experiments. The resulting
bounds would 
be larger than those shown in the two figures.
However, the main points of the discussion would remain unchanged.

Fig.~\ref{nnng1} illustrates the resulting constraints on a 1.5 TeV
$Z'$ at a 500 GeV
collider for different
observables and experimental parameters, including luminosity and 
beam polarization. 
We see that we can get some interesting constraints
even though
the $Z'$ is considerably heavier than the centre-of-mass energy.
The region which cannot be resolved by the observables is between the 
two corresponding lines and contains the couplings of the SM. Hence the
star in this figure corresponds to the SSM($Z'$).
For the cases where only one bounding line is shown, the second line
is outside the figure.
$R_\nu(Z')$ and $L_\nu(Z')$ are mainly constrained by the interference of the 
$Z'$ exchange with the SM. The strongest constraint is on 
the $Z'$ coupling to left-handed neutrinos.
This makes the constraints especially simple.

First we consider an integrated luminosity of 500 fb$^{-1}$. The total
unpolarized cross
section gives the strongest constraint.
The constraints from energy and angular distributions (with 
10 equal size bins) were also considered but
they give no improvement.
The constraint from $A_{LR}$ is shown for two polarization cases: 90\%
electron 
beam polarization and the case of a
collider with
a $P^-=90\%$ polarized electron beam and a $P^+=60\%$ polarized positron 
beam.
Even for the latter case, the constraint from $A_{LR}$ is worse
than that from the total cross section.
We mention here for completeness that two polarized beams give not only
a high effective polarization but also
a small effective polarization error \cite{marciano}.

The constraint obtained with an 
integrated luminosity of $L_{\rm int}=50$ fb$^{-1}$ is also
shown in Fig.~\ref{nnng1}, to contrast with the high luminosity case. We
see that for $L_{\rm int}=500$ fb$^{-1}$
a systematic error of 1\% relaxes the constraints considerably and 
dilutes the advantage of high luminosity.
Thus, both small systematic errors and a high luminosity collider 
are highly desired for the proposed measurement.

Fig.~\ref{nnng2} shows the possible constraints on  $R_\nu(Z')$ and 
$L_\nu(Z')$ from $\sigma$ and $A_{LR}$,
including systematic errors, for two representative 
$Z'$ masses, 0.75 TeV and 1 TeV.
The constraints become much stronger as the $Z'$ mass is decreased.
So far, we assumed that the $Z'e^+e^-$ couplings,  $R_e(Z')$ and $L_e(Z')$, 
are precisely known.
However, they must be measured (with errors) by another experiment.
Fig.~4(b) of \cite{lmu0296} illustrates such a measurement 
for a collider with a luminosity of 20 fb$^{-1}$.
To estimate their influence on the  $R_\nu(Z')$, $L_\nu(Z')$ constraint,
we make use of the errors on $R_e(Z')$ and $L_e(Z')$ given in 
\cite{lmu0296}. 
Our input for the errors of the $Z'e^+e^-$ couplings for
$M_{Z'}=1.0$ TeV and
$0.75$ TeV are obtained from those for $1.5$ TeV by the scaling relation
(2.63) in \cite{physrep}.
We see that 
the uncertain knowledge of the $Z'e^+e^-$ couplings leads to
only 
slightly weaker constraints on $R_\nu(Z')$ and $L_\nu(Z')$. 
However, Fig.~\ref{nnng2} shows that this effect is only important 
for a relatively heavy
$Z'$ and for $R_\nu(Z')$ (even at lower $Z'$ masses) for
which the constraints are already weak.

Finally, we mention that there is no sign ambiguity in the measurement 
of  $R_\nu(Z')$ and $L_\nu(Z')$ if the signs of the  $Z'e^+e^-$ couplings 
are known.
It was noted \cite{lmu0296} that the $Z'e^+e^-$ couplings have a two-fold sign
ambiguity if measured in the process $e^+e^-\rightarrow e^+e^-$ alone.
If this ambiguity exists, it induces a related sign ambiguity for 
$R_\nu(Z')$ and $L_\nu(Z')$.
If the sign ambiguity in the $Z'e^+e^-$ couplings is resolved
\cite{physrep}  
(i.e. by measurements obtained from the 
process $e^+e^-\rightarrow W^+W^-$ below the
$Z'$ resonance or by measurements at the $Z'$ resonance) it also disappears 
in our constraints on $R_\nu(Z')$ and $L_\nu(Z')$.

In Figs.~\ref{nnng3} to \ref{nnng6}, we shall 
assume that there is no signal from a 
$Z'$ but that a 
signal from a $W'$ has been observed.
This could happen in models where the $W'$ is considerably 
lighter than the $Z'$. We recognize that this
particular scenario is unlikely in the context of the models we consider.
For
instance, in the UUM, the $W'$ and $Z'$ masses are approximately
equal and
there would most likely be a signal observed for the $Z'$ in addition to
the
$W'$. The situation is similar in the LRM, where the relationship
between the $W'$ and $Z'$ masses is given in Eq.~(\ref{massrel}). Thus, it
should be
understood that our results for the case of a
$W'$ only
represent an estimate of the reach of this process in constraining $W'$
couplings, rather than precision limits in the context of a full
understanding of the physics realized in nature. We use this simple
scenario
in order to indicate sensitivity to various parameters, such as the
observables used and the luminosity.
Alternatively, a known $Z'$ could be included in the following analysis.
Again, the experimental errors on the measured $Z'$ parameters
would enlarge the errors of the $W'$ measurements but not change the
main conclusions.
We will see that the process $e^+e^-\rightarrow\nu\bar\nu\gamma$ 
can give model independent 
constraints on the quantities $L_l(W')$ and $R_l(W')$ 
for $W'$ masses considerably larger than the center-of-mass energy.
We only probe $l=e$ directly, but we are assuming lepton universality
throughout.

Fig.~\ref{nnng3} is similar to Fig.~\ref{nnng1}, but 
it shows the constraints on the $W'$
couplings. In this figure,
for illustration, we assume there exists a $W'$ with 
SM couplings but with a mass of 1.5 TeV and that the right-handed
neutrino is light enough to be produced.
We find that the left- and, to some extent, the right-handed $W'$ coupling 
can be constrained. The figure illustrates the use of different
combinations of 
$\sigma$ and $A_{LR}$, and of different beam polarizations.
The unpolarized cross section  mainly constrains  the left-handed
$W'$ coupling because left-handed electrons give its dominant contribution.
The constraints from energy and angular distributions give 
almost no improvement for the model considered here.
The constraint from $A_{LR}$ is complementary to that from $\sigma$.
It is shown for the two cases of 90\% electron beam polarization and for
90\% electron beam polarization with 60\% positron polarization.
We see that $\sigma$ and $A_{LR}$ together give the 
best constraints on the couplings.

The constraints on the $W'$ couplings have a two-fold sign ambiguity;
 nothing is changed by a simultaneous change of the sign of 
$L_l(W')$ and $R_l(W')$.
The reason for this ambiguity lies in the squared amplitude, Eq.\
(\ref{m2}), where these couplings
always enter as squares or as a product of left and right $W'$ couplings.
In the case where we have only a weak $W'$ signal, the two regions allowed
by this ambiguity 
overlap into one large region.

In Fig.~\ref{nnng4}, we show constraints on the $W'$ couplings from
$\sigma$
 and 
$A_{LR}$ combined. In this figure, we illustrate the use of
different luminosities and the inclusion of a systematic error.
We have the same two well separated 
regions for the case of high luminosity and no systematic error
as in Fig.~\ref{nnng3}. 
These two regions become larger for low luminosity and 
no systematic error.
We are left with one large region after the inclusion of a systematic error 
of 2\% for $\sigma$ and 1\% for $A_{LR}$.
As in the case of extra neutral gauge bosons, 
small systematic errors
{\it and} high luminosity are necessary for a coupling measurement.

In Fig.~\ref{nnng5}, we show how the constraints on the $W'$ couplings vary 
for different $W'$ masses.
The constraint for $M_{W'}=1.5$ TeV is identical to that from 
Fig.~\ref{nnng4}.
We see that the constraint on the $W'$ couplings improves dramatically 
for lower $W'$ masses.

Fig.~\ref{nnng6} illustrates the possibility of
discrimination between different models.
We see that a $W'$ with SM couplings ($W'_L$) can be separated from the SM.
A $W'$ with pure right-handed couplings ($W'_R$) with a strength of the 
left-handed coupling of the SM $W$ cannot be distinguished from the SM case. 

Looking at the squared amplitude, Eq.\ (\ref{m2}), 
we see that the constraints shown in Figs.~\ref{nnng3} to \ref{nnng6} are, 
to a good approximation,
valid for the combinations $L_l(W')/M_{W'}$ and  $R_l(W')/M_{W'}$, and not
for the couplings and the mass separately.
We have fixed the $W'$ mass here for illustrational purposes.
If a $W'$ is found with a mass different from our assumptions, the constraint
on its couplings can be found by the appropriate scaling of our 
results.

So far, we considered model independent bounds on the couplings of a
single extra
gauge boson while neglecting the existence of other extra gauge bosons.
However, typically, extra neutral and charged gauge bosons 
simultaneously influence the observables. We consider this situation for
the LRM and the UUM.

In Fig.~\ref{nnng7}, we consider the Left-Right symmetric model.
For $M_{W'}=0.75$ TeV,  Eq.~(\ref{massrel})
gives $M_{Z'}=0.90(1.27)$ TeV for $\kappa=1$ and $\rho=1(2)$.
We show the constraints on the couplings of the $W'$ for $\rho=1$ 
obtained by two different fitting strategies.
First, we ignore the $Z'$ completely, and second, we take the $Z'$ into
account
assuming exact knowledge of its couplings.
We see that the two curves are quite close.
The reason is that our process is not very sensitive to such a $Z'$.
These two curves  are very similar to those for the
$W_R$ 
and the SM in Fig.~\ref{nnng6} because we are not very sensitive to
a right-handed $W'$.
The case of $\rho=2$ predicts a heavier $Z'$, which
produces constraints differing even less from each other than those 
for $\rho=1$,
so we do not show them.
To demonstrate how the constraints change for a larger signal,
we repeated the same procedure 
with $M_{W'}=550$ GeV. This number (and the mass of the associated $Z'$)
are at the edge of the present exclusion limit \cite{pdb}.
Although the constraints improve a bit, they are still not very impressive.

Fig.~\ref{nnng8} is similar to Fig.~\ref{nnng7} 
but here we consider the Un-Unified model.
We examine the cases $M_{W'}=M_{Z'}=0.75$ TeV 
and 
$M_{W'}=M_{Z'}=0.55$ TeV. 
We show the constraints on the couplings of the $W'$ 
obtained using the same 
two fitting strategies described for Fig.~\ref{nnng7}.
Even for masses of $0.75$ TeV, the two curves are better
separated than in LRM.
For masses of 0.55 TeV, the wrong fitting strategy gives a region which
is outside the true $W'$ coupling.
This shows that such a light $Z'$ cannot be 
ignored in the fitting procedure.

The process $e^+e^-\rightarrow f\bar f$ and searches in hadron collisions
are
more sensitive to $Z'$ discovery than $e^+e^-\rightarrow\nu\bar\nu\gamma$.
A $Z'$ signal will always be detected in the cases where the
$Z'$ contribution
is relevant for a $W'$ constraint from $e^+e^-\rightarrow\nu\bar\nu\gamma$.
This information
from other experiments will be required for a reliable $W'$ constraint
from $e^+e^-\rightarrow\nu\bar\nu\gamma$.

\section{Conclusions}

In this paper, we studied the sensitivity of the process $e^+e^-\to 
\nu\bar{\nu}\gamma$ to extra gauge bosons.  We used this process to 
find discovery limits and to see how well one could measure 
the couplings of extra gauge bosons that are expected in 
extensions of the standard model.  

For the discovery limits we 
focused on $W'$'s since one can put better limits on $Z'$'s from 
other processes, such as $e^+e^-\to f\bar{f}$, while, on the other 
hand,  no similar limits exist on $W'$'s. The highest reach was 
obtained by binning the $d\sigma/dE_\gamma $ distribution although 
comparable results were obtained using the total cross section after 
the $Z$ radiative return was eliminated. The discovery reach is 
typically in the 1-6~TeV range depending on the specific model, the 
center of mass energy,  and the 
assumed integrated luminosity.  These results are substantially 
degraded if one includes systematic errors.  For the $W_R$ boson, for 
which LHC discovery limits are available, the discovery limits are,
for $g_R=g_L$, 
$M_{W'}$= 1.2, 1.6, and 1.9 TeV for $\sqrt{s}=$ 500, 1000, and 1500 
GeV respectively assuming $L_{\rm int}=500$ fb$^{-1}$ relative to a reach of 
$~5.9$~TeV at the LHC.

Although the discovery reach for $W'$'s 
of this process is not competitive with the reach of the LHC, precision 
measurements can give information on extra gauge boson couplings 
which complements the LHC.  In particular, if the LHC were to discover 
a $Z'$ or $W'$ the process $e^+e^- \to \nu\bar{\nu}\gamma$ could 
constrain $Z'$ and $W'$ couplings. For a $Z'$, this would be the best 
measurement of the $Z' \nu \bar{\nu}$ couplings.  
For $W'$ couplings, reliable measurements would require information 
from, for example, 
$e^+e^-\rightarrow f\bar f$ and searches in hadron collisions which 
would always detect a $Z'$ signal in the cases where its contribution
is relevant for a $W'$ constraint by $e^+e^-\rightarrow\nu\bar\nu\gamma$.
Finally, we emphasize that 
to make measurements of the extra gauge boson couplings, high 
luminosity will be needed and it will be very
important to reduce the systematic uncertainties as much as possible.

\acknowledgments

This research was supported in part by the Natural Sciences and Engineering 
Research Council of Canada.
S.G., P.K., and B.K.\ thank Dean Karlen and A.L. thanks Graham Wilson 
for useful discussions.

\section*{appendix}
Here we give explicit parametrizations of the momenta defined in the
frame where $q_-, q_+$ are back-to-back and
$p_+$ defines the $\hat{z}$ axis, suitable for use with 
the phase-space (\ref{phasesp}):
\begin{eqnarray}
\nonumber
p_+ &=& (\omega_+;0,0,\omega_+) \\
\nonumber
p_- &=& (\omega_-; \omega_k\sin\psi, 0, \omega_k\cos\psi - \omega_+) \\
\nonumber
k &=& (\omega_k; \omega_k\sin\psi,0,\omega_k\cos\psi) \\
\nonumber
q_+ &=& (\omega'_+; \omega'_+\sin\theta\cos\varphi, 
\omega'_+\sin\theta\sin\varphi,\omega'_+\cos\theta) \\
q_- &=& (\omega'_+; -\omega'_+\sin\theta\cos\varphi, 
-\omega'_+\sin\theta\sin\varphi,-\omega'_+\cos\theta),
\end{eqnarray}
where
\begin{eqnarray}
\nonumber
\omega_- &=& \frac{s-k_-}{2\sqrt{s'}}, \,\,\,\,
\omega_+ = \frac{s-k_+}{2\sqrt{s'}}, \,\,\,\,
\omega'_+ = \frac{\sqrt{s'}}{2}, \\
\omega_k &=& \frac{s-s'}{2\sqrt{s'}}, \,\,\,\,
\cos\psi = \frac{sk_- - s'k_+}{(s-k_+)(s-s')}.
\end{eqnarray}
It is arbitrary whether $\sin\psi$ is taken as positive or negative
as long as one is consistent.

\newpage
\begin{figure}
\centerline{\epsfig{file=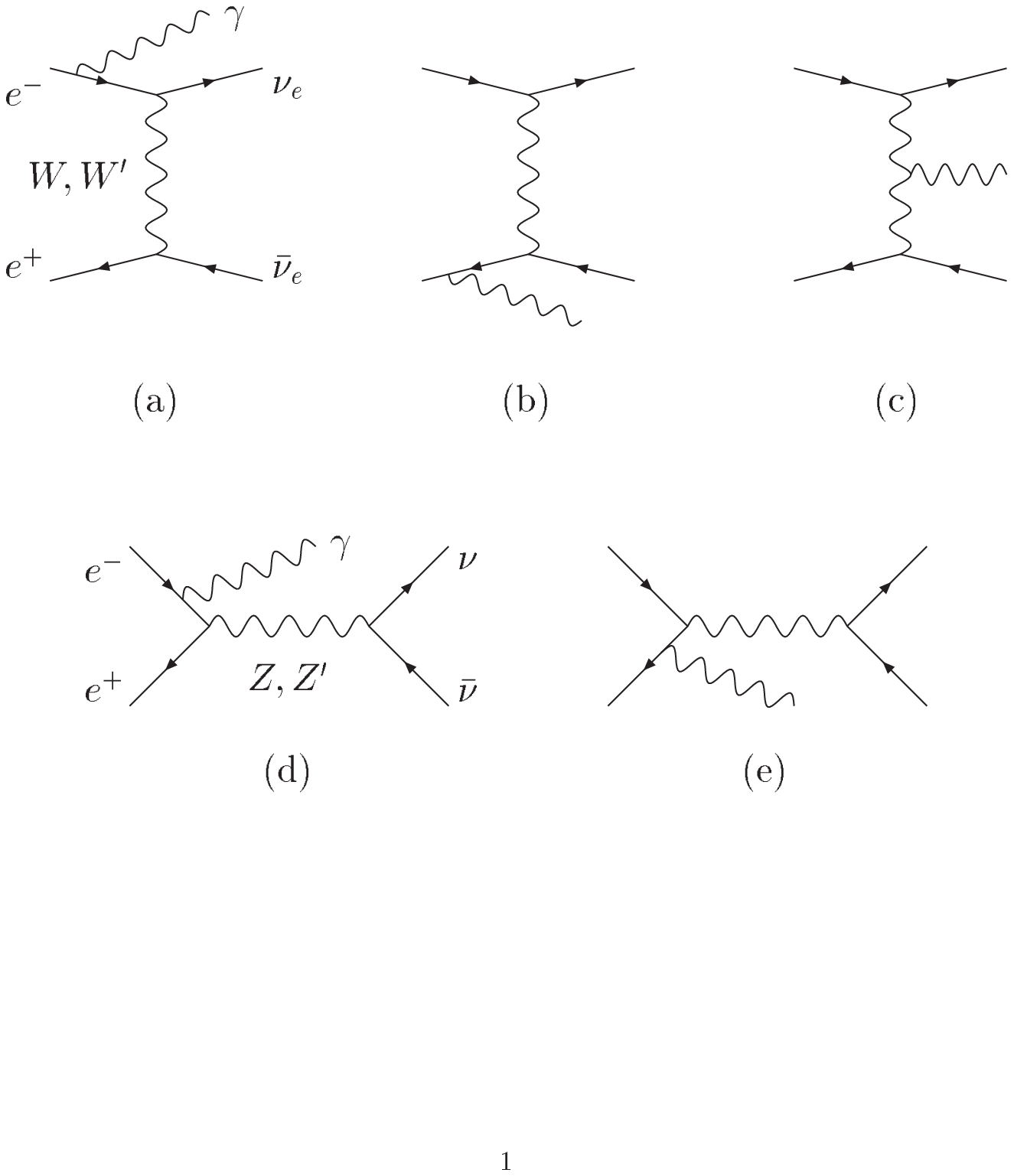,width=6.5in}}
\vspace{20pt}
\caption{The Feynman diagrams contributing to the process
$e^+e^-\rightarrow\nu\bar{\nu}\gamma$ in leading order.}
\label{Fig1}
\end{figure}
\newpage
\begin{figure}
\vspace{-1.3in}
\centerline{\epsfig{file=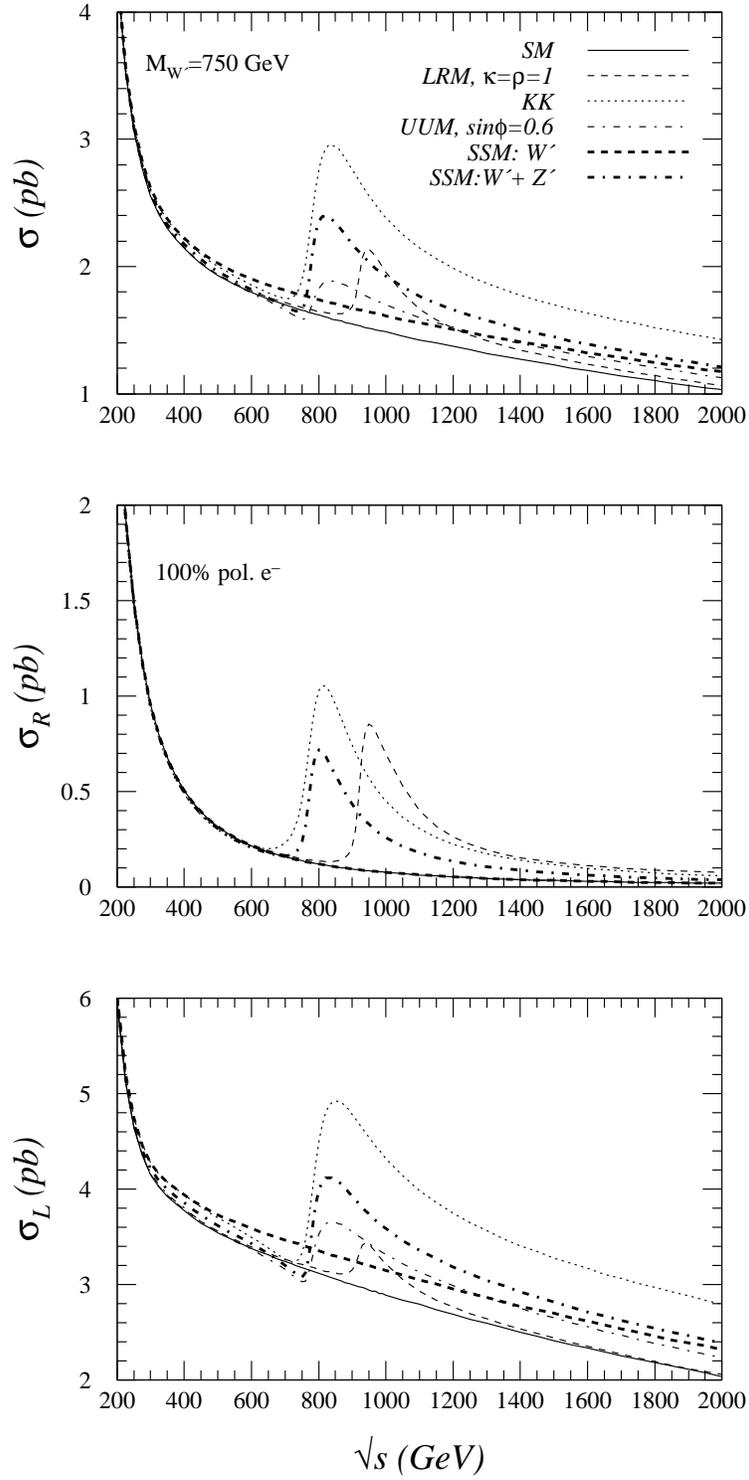,width=5in}}
\vspace{-0.5in}
\caption{The total cross sections $\sigma$, $\sigma_L$ and $\sigma_R$
versus $\sqrt{s}$ for $M_{W'}=750$ GeV. For $\sigma_L$ and $\sigma_R$,
100\% $e^-$ polarization is used. Results are given for the SM 
(solid line), LRM (dashed line), KK model (dotted line), UUM
(dash-dotted line), SSM($W'$) (thick dashed line) and SSM($W'+Z'$)
(thick dash-dotted line).
}
\label{Fig2}
\end{figure}
\newpage
\begin{figure}
\vspace{-1in}
\centerline{\epsfig{file=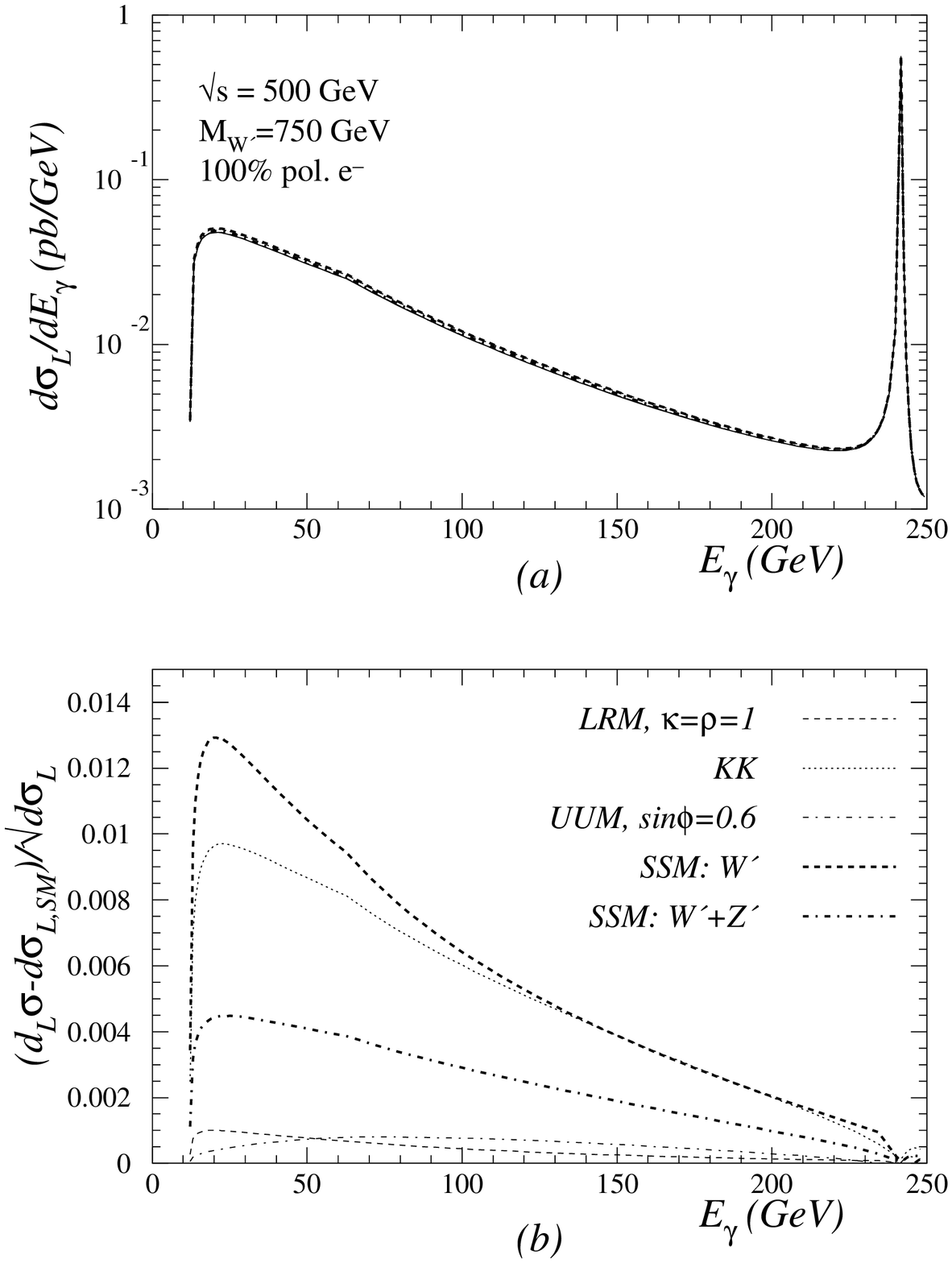,width=6.5in}}
\caption{(a) Left-handed differential cross section versus energy;
(b) relative statistical significance of the deviation from the SM,
for $\sqrt{s}=500$ GeV and $M_{W'}=750$ GeV. 100\% $e^-$ polarization
is used. Lines as in Fig.\ 2.
}
\label{Fig3}
\end{figure}
\newpage
\begin{figure}
\vspace{-1in}
\centerline{\epsfig{file=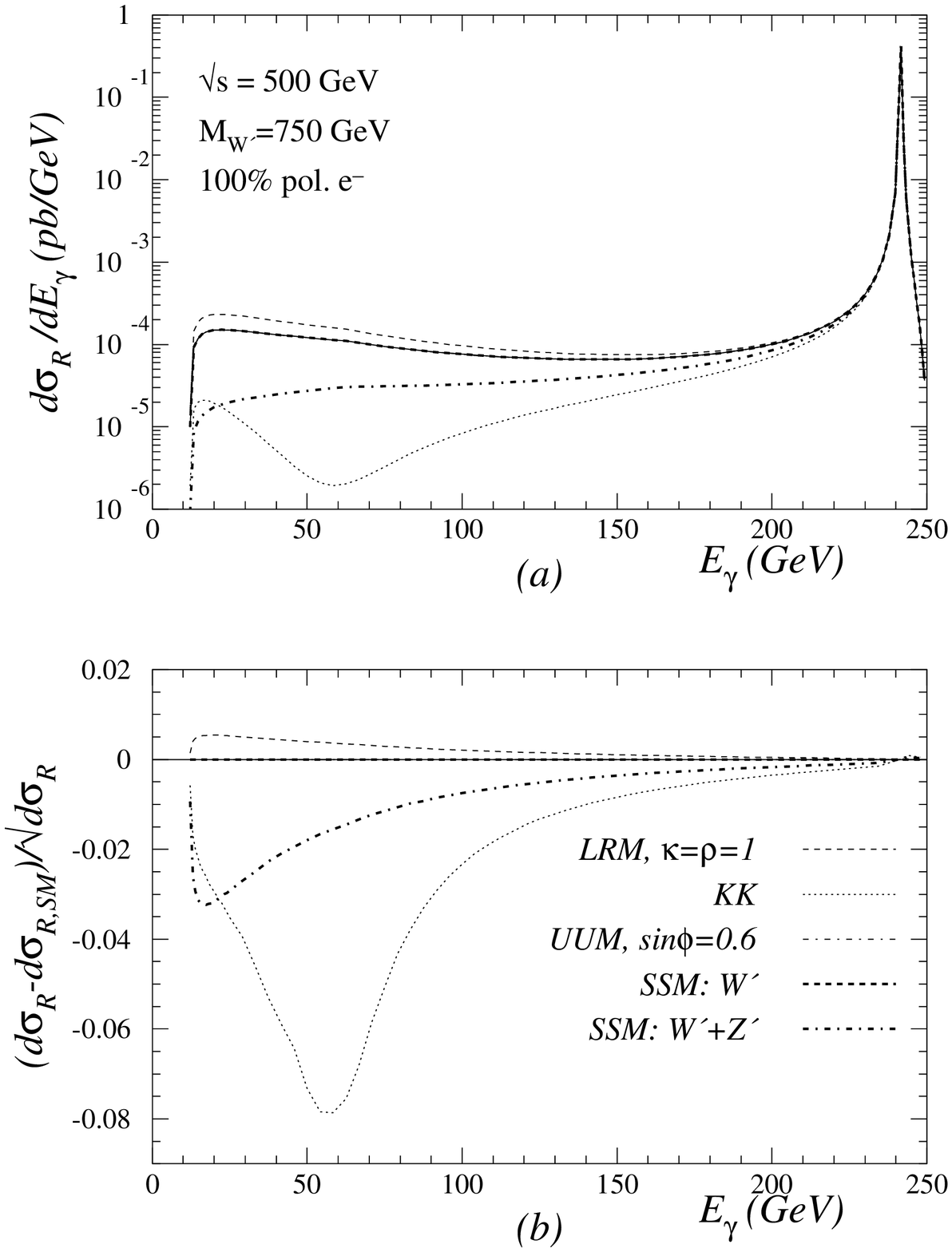,width=6.5in}}
\caption{(a) Right-handed differential cross section versus energy;
(b) relative statistical significance of the deviation from the SM,
for $\sqrt{s}=500$ GeV and $M_{W'}=750$ GeV. 100\% $e^-$ polarization
is used. Lines as in Fig.\ 2.
}
\label{Fig4}
\end{figure}
\newpage
\begin{figure}
\vspace{-1in}
\centerline{\epsfig{file=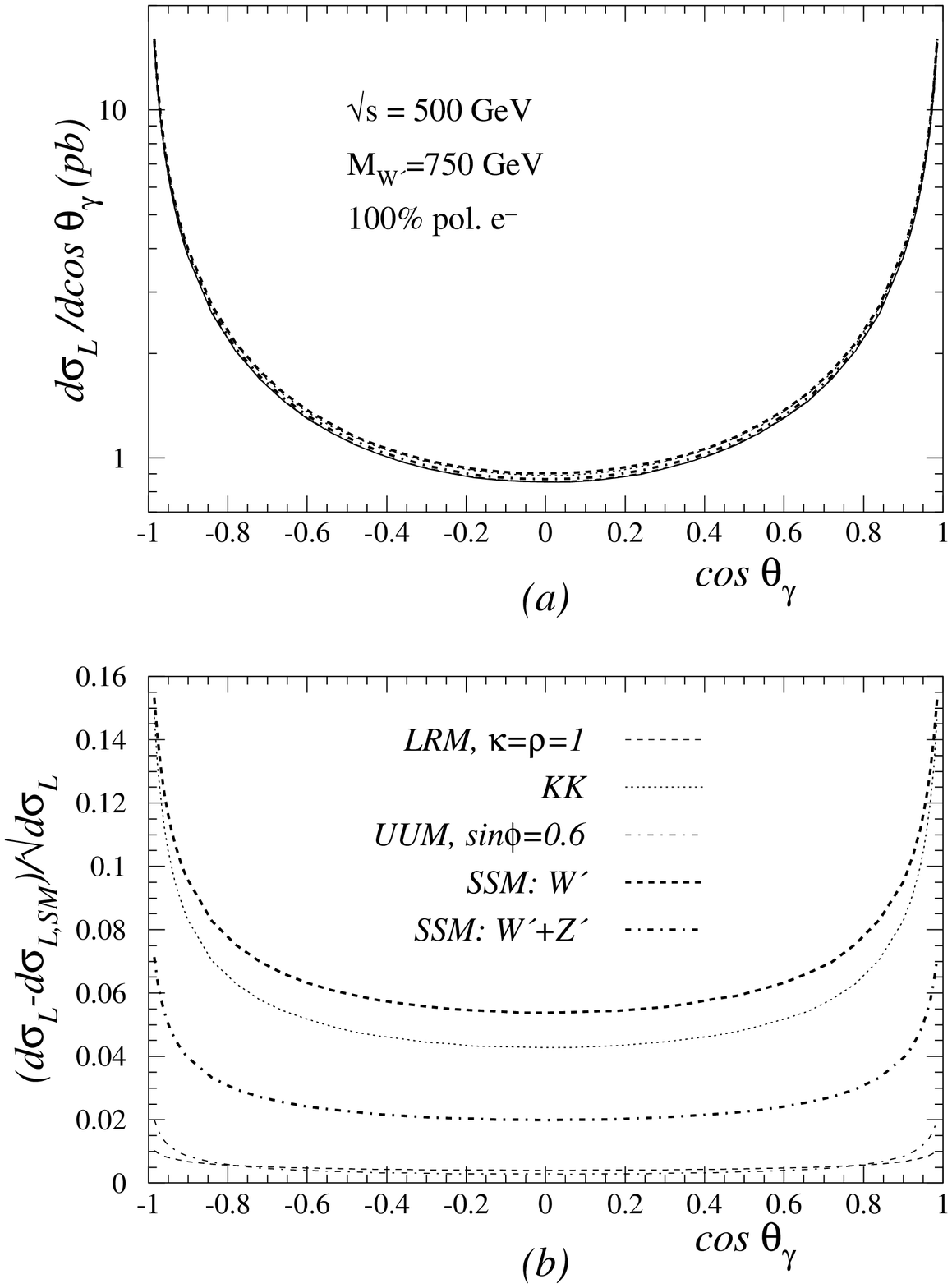,width=6.5in}}
\caption{(a) Left-handed differential cross section versus $\cos\theta_\gamma$;
(b) relative statistical significance of the deviation from the SM,
for $\sqrt{s}=500$ GeV and $M_{W'}=750$ GeV. 100\% $e^-$ polarization
is used. Lines as in Fig.\ 2.
}
\label{Fig5}
\end{figure}
\newpage
\begin{figure}
\vspace{-1in}
\centerline{\epsfig{file=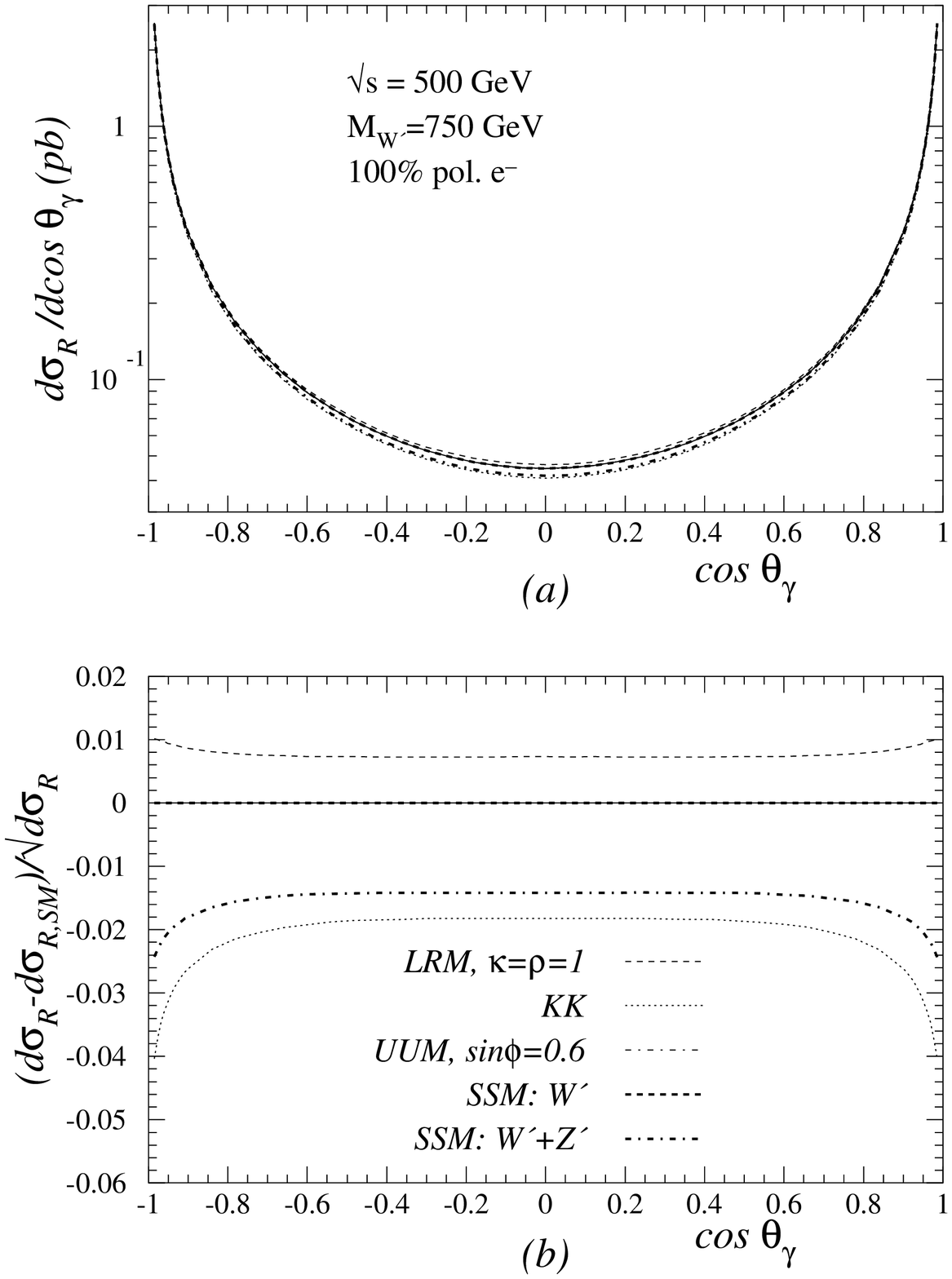,width=6.5in}}
\caption{(a) Right-handed differential cross section versus $\cos\theta_\gamma$;
(b) relative statistical significance of the deviation from the SM,
for $\sqrt{s}=500$ GeV and $M_{W'}=750$ GeV. 100\% $e^-$ polarization
is used. Lines as in Fig.\ 2.
}
\label{Fig6}
\end{figure}
\newpage
\begin{figure}
\vspace{-1in}
\centerline{\epsfig{file=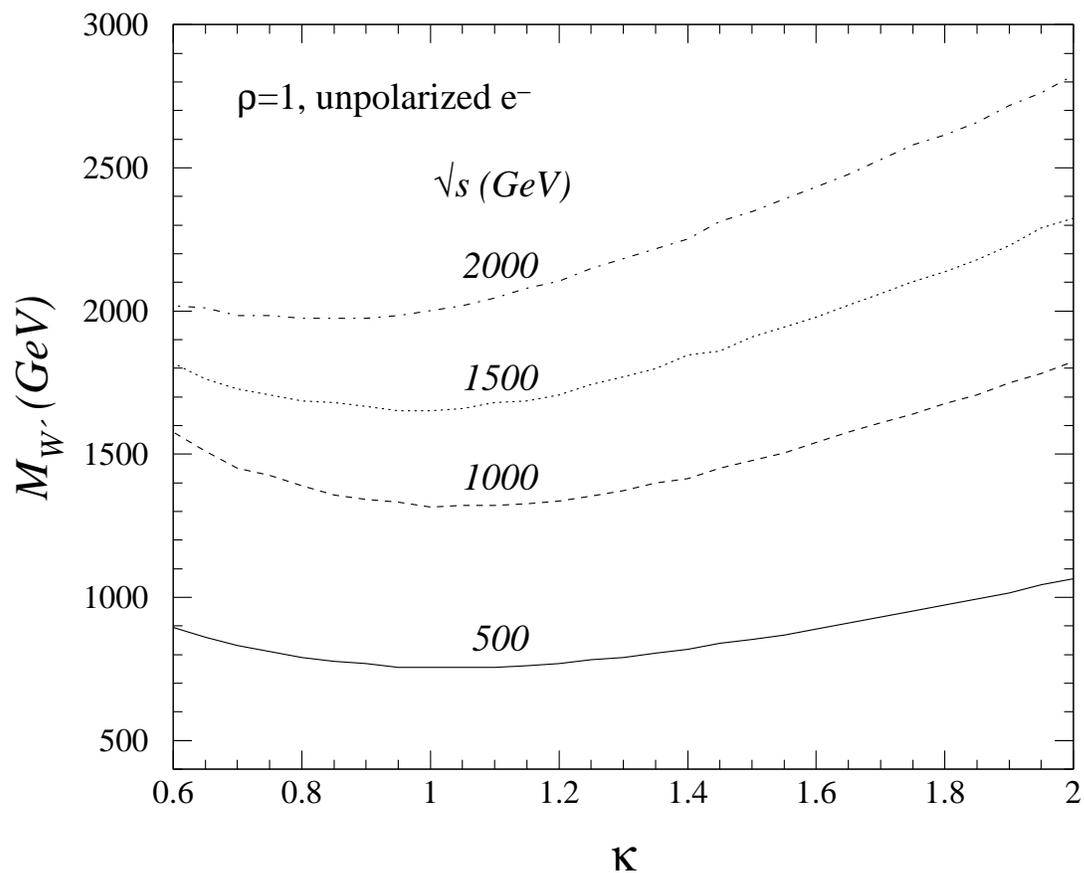,width=6.5in}}
\caption{LRM ($\rho=1$) unpolarized 95\% C.L. $W'$ mass limits versus 
$\kappa$, obtained
for $\sqrt{s}=500$, 1000, 1500 and 2000 GeV using 
$d\sigma/dE_\gamma$ as the observable and an integrated luminosity
of 50 fb$^{-1}$ for $\sqrt{s}=500$ GeV and  200 fb$^{-1}$ for
the higher energies.
Only statistical errors are used.
}
\label{Fig7}
\end{figure}
\newpage
\begin{figure}
\vspace{-1in}
\centerline{\epsfig{file=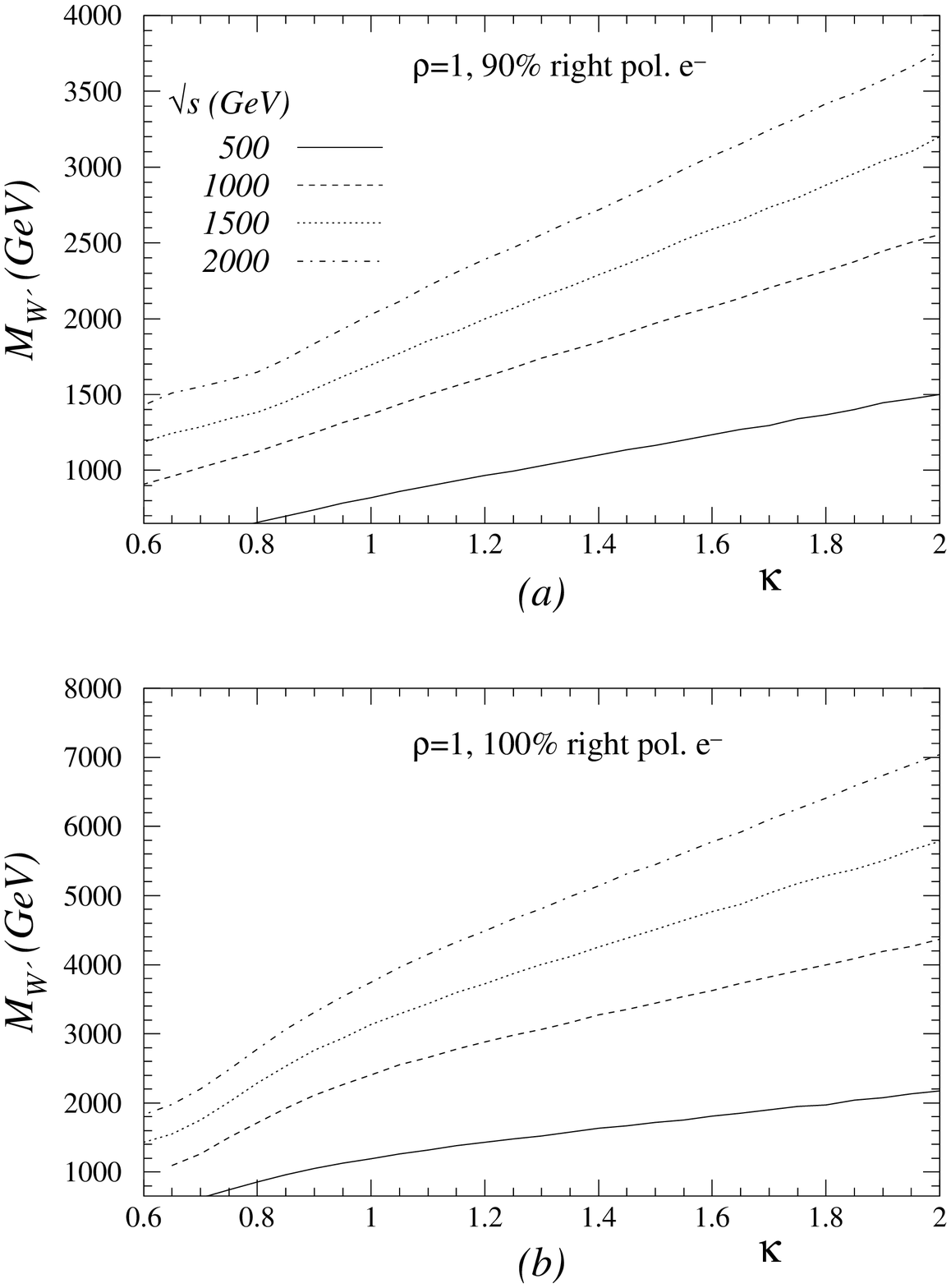,width=6.5in}}
\caption{As Fig.\ 7, except (a) for 90\% right-polarized electrons,
(b) for  100\% right-polarized electrons.
}
\label{Fig8}
\end{figure}
\newpage
\begin{figure}
\vspace{-1in}
\centerline{\epsfig{file=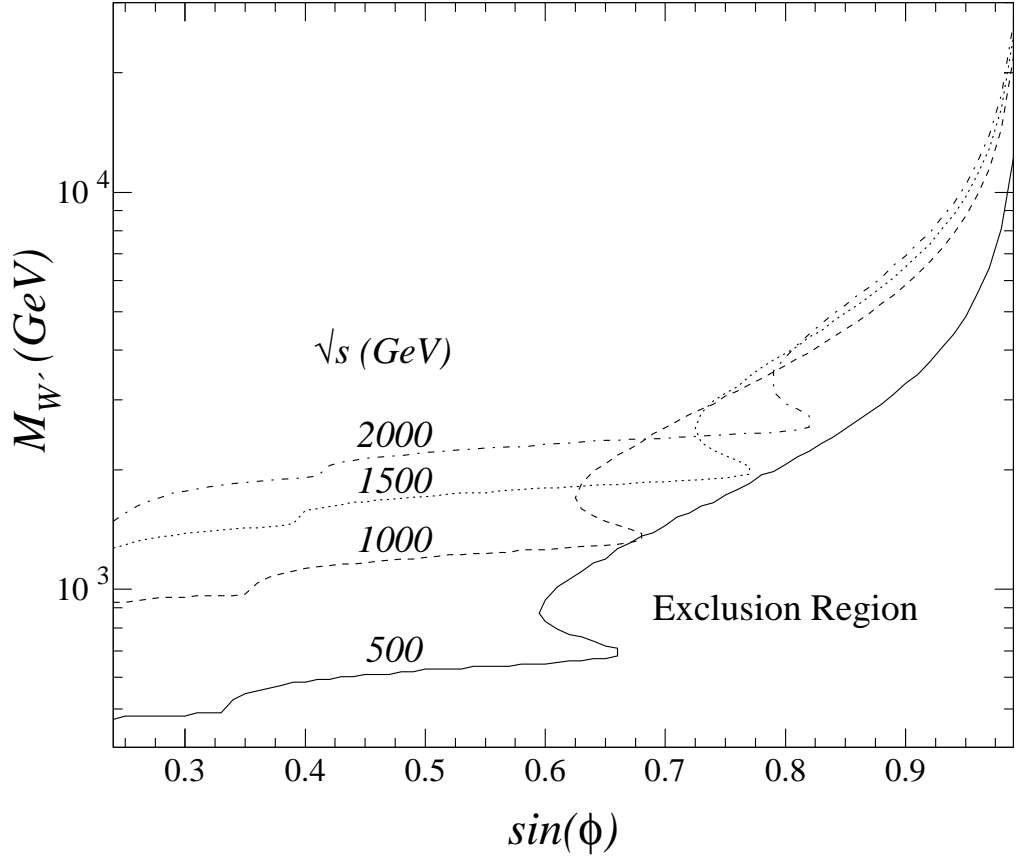,width=6.5in}}
\caption{UUM unpolarized 95\% C.L. $W'$ mass limits versus $\sin\phi$,
obtained for $\sqrt{s}=500$, 1000, 1500 and 2000 GeV using 
$d\sigma/dE_\gamma$ as the observable and an integrated luminosity
of 50 fb$^{-1}$ for $\sqrt{s}=500$ GeV and  200 fb$^{-1}$ for
the higher energies. Only statistical errors are used. The region to the
right of the various curves is the region which may be excluded by
experiment.
}
\label{Fig9}
\end{figure}
%

\newpage
\begin{figure}
\vspace{-0.5in}
\mbox{
\epsfysize=5in
\epsffile[0 0 500 500]{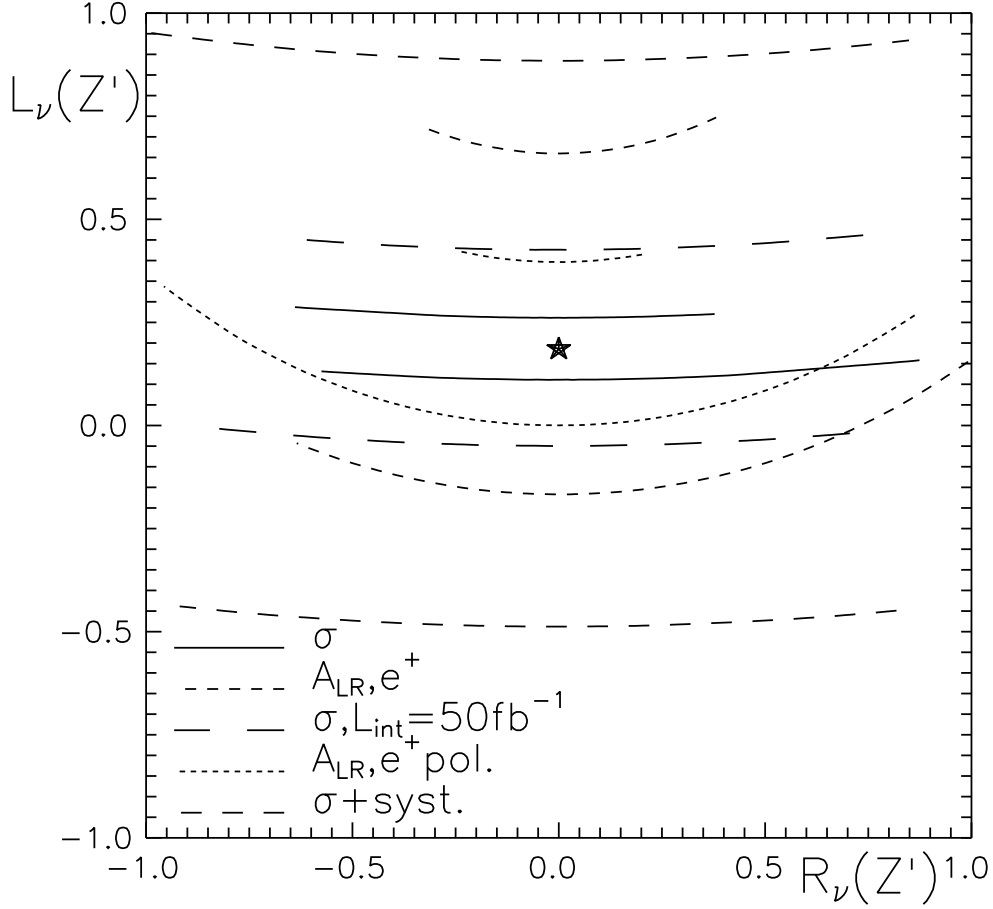}
}
\vspace{0.5in}
\caption{Constraints on the $Z'\nu\bar\nu$ couplings $R_\nu(Z')$ 
and $L_\nu(Z')$ 
below the $Z'$ peak using different observables. We take
$\sqrt{s} =0.5$ TeV, $\ M_{Z'}=1.5$ TeV and
$L_{\rm int}=500$ fb$^{-1}$, except in the indicated case 
where it is $50$ fb$^{-1}$.
The polarization of the electron beam is 90\% and
the positron beam is unpolarized, except in the indicated case where it is 
60\% polarized.
Only statistical errors are included in this figure, except in the indicated 
case where a systematic error of 1\% is included for $\sigma$.
The assumed model is a $Z'$ in the Sequential Standard Model [SSM ($Z'$)],
indicated by a star.
}
\label{nnng1}
\end{figure}
\newpage
\begin{figure}
\vspace{-0.5in}
\mbox{
\epsfysize=5in
\epsffile[0 0 500 500]{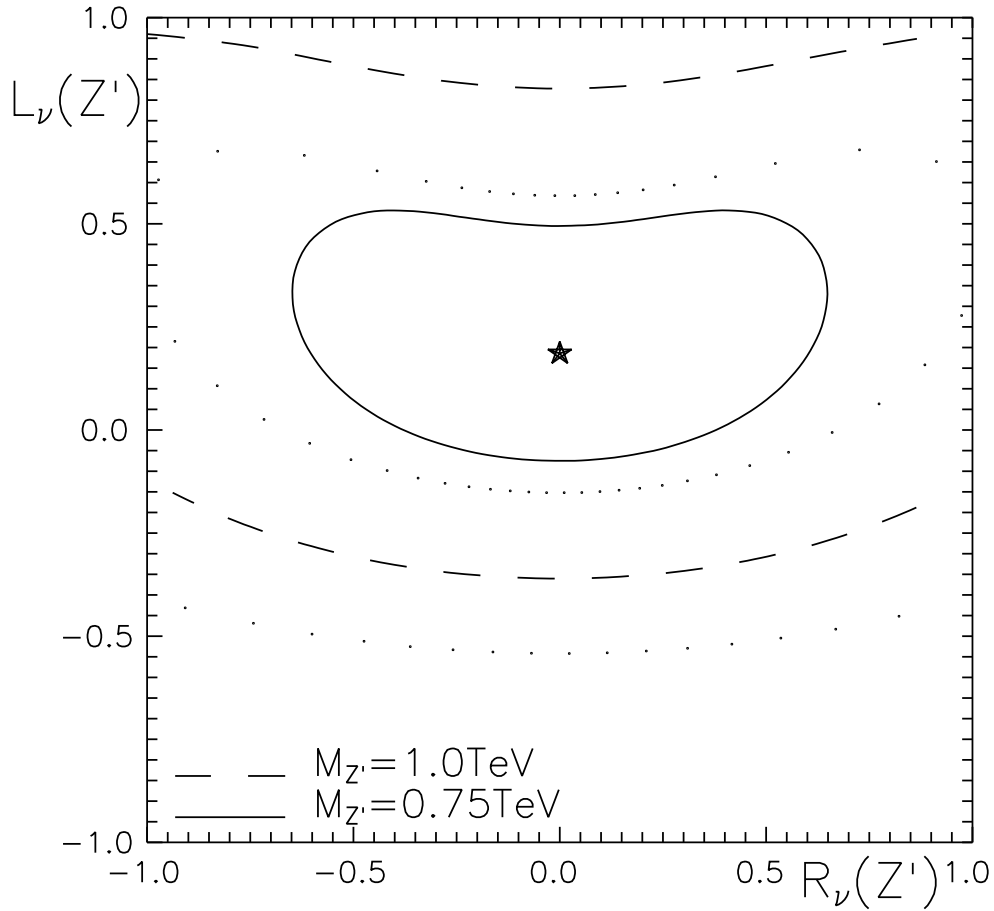}
}
\vspace{0.5in}
\caption{Constraints on  $R_\nu(Z')$ and $L_\nu(Z')$ below the $Z'$ peak using
$\sigma$ and $A_{LR}$ combined as observables.
The lines show the results for two different $Z'$ masses.
The dots indicate how the constraints relax if the error on the $Z'e^+e^-$ 
coupling measurement is included as described in the text. 
We take
$\sqrt{s}=0.5$ TeV, $L_{\rm int}=500$ fb$^{-1}$ and a 
systematic error of 2\% (1\%) for $\sigma$ ($A_{LR}$).
The assumed model [SSM ($Z'$)] is indicated by a star.
}
\label{nnng2}
\end{figure}
\newpage
\begin{figure}
\vspace{-0.5in}
\mbox{
\epsfysize=5in
\epsffile[0 0 500 500]{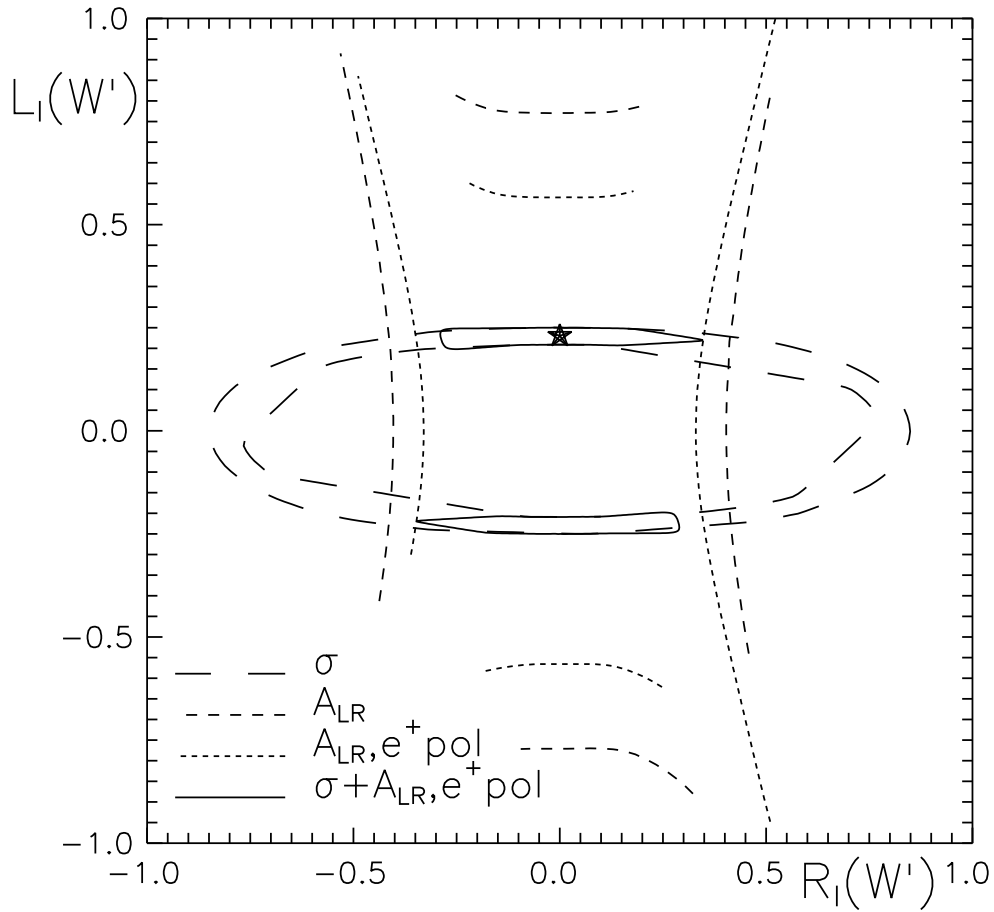}
}
\vspace{0.5in}
\caption{Constraints on the $W'$ couplings using $\sigma$, $A_{LR}$ and 
using $\sigma$ and $A_{LR}$ combined as observables. We take
$\sqrt{s}=0.5$ TeV, $L_{\rm int}=500$ fb$^{-1}$ and $M_{W'}=1.5$ TeV.
Only statistical errors are included in this figure.
90\% electron and, where indicated, 60\% positron polarization are used.
The assumed model [SSM ($W'$)] is indicated by a star.
}
\label{nnng3}
\end{figure}
\newpage
\begin{figure}
\vspace{-0.5in}
\mbox{
\epsfysize=5in
\epsffile[0 0 500 500]{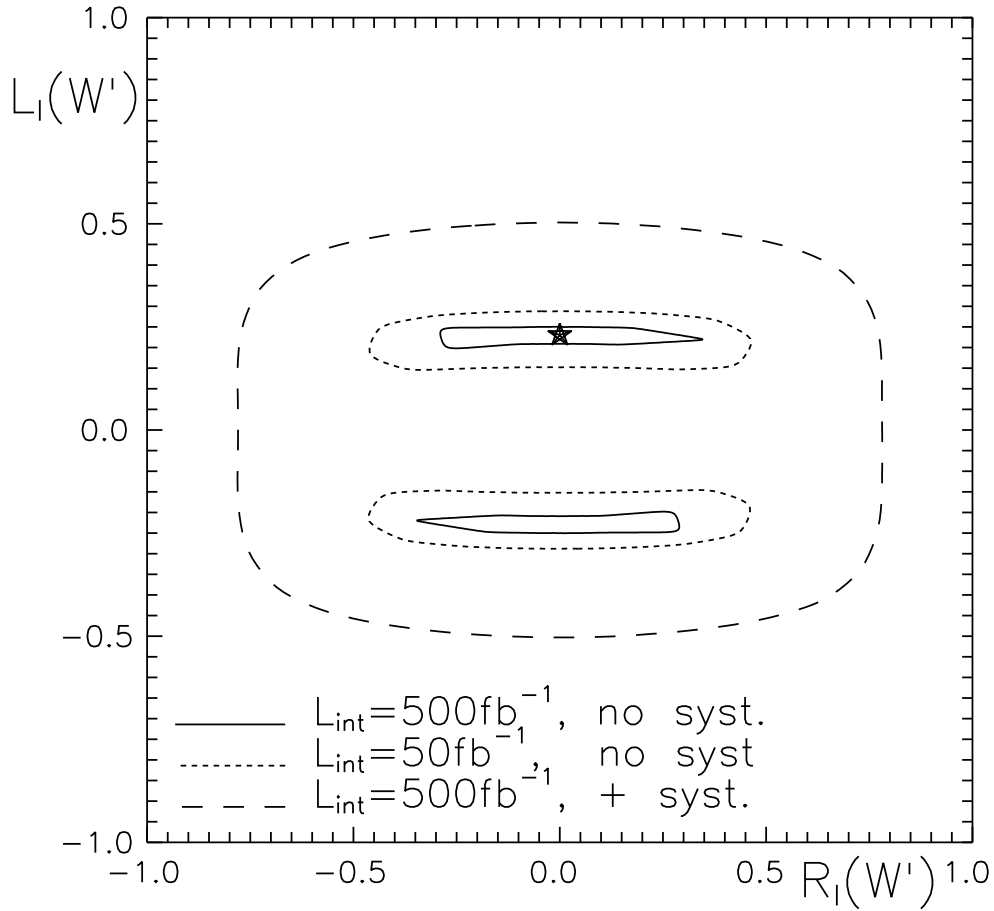}
}
\vspace{0.5in}
\caption{Constraints on the $W'$ couplings using $\sigma$ and $A_{LR}$ combined.
90\% electron and 60\% positron polarization are used. We take
$\sqrt{s}=0.5$ TeV,  $M_{W'}=1.5$ TeV and 
$L_{\rm int}=500$ fb$^{-1}$, except in the 
indicated case where it is $50$ fb$^{-1}$.
Only statistical errors are used, except in the indicated case where a 
systematic error of 2\% (1\%) is included for $\sigma$ ($A_{LR}$).
The assumed model [SSM ($W'$)] is indicated by a star.
}
\label{nnng4}
\end{figure}

\newpage
\begin{figure}
\vspace{-0.5in}
\mbox{
\epsfysize=5in
\epsffile[0 0 500 500]{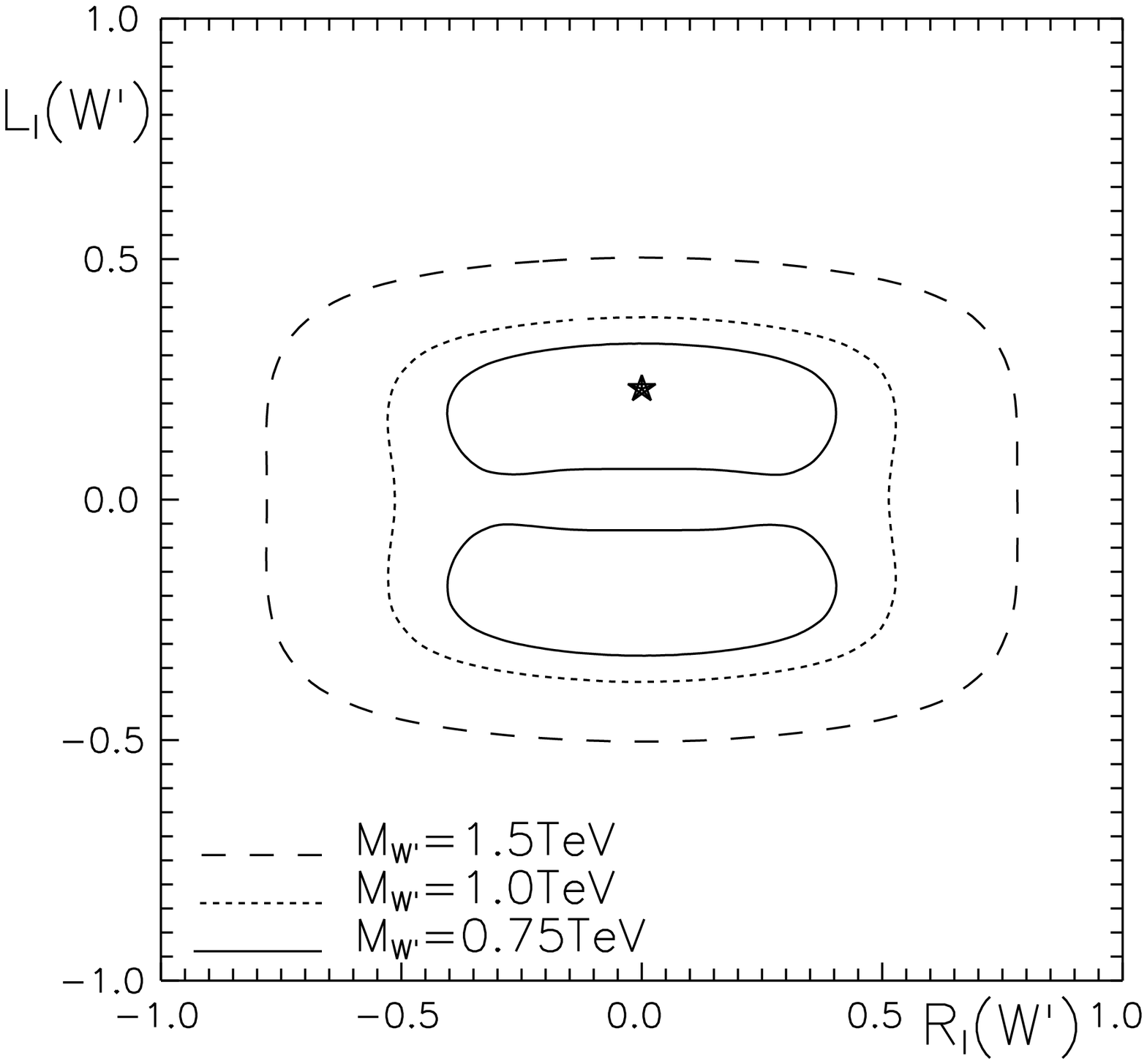}
}
\vspace{0.5in}
\caption{Constraints on the $W'$ couplings using $\sigma$ and $A_{LR}$ combined
for different $W'$ masses. We take
90\% electron and 60\% positron polarization,
$\sqrt{s}=0.5$ TeV and $L_{\rm int}=500$ fb$^{-1}$.
A systematic error of 2\% (1\%) is included for $\sigma$ ($A_{LR}$).
The assumed model [SSM ($W'$)] is indicated by a star.
}
\label{nnng5}
\end{figure}
\newpage
\begin{figure}
\vspace{-0.5in}
\mbox{
\epsfysize=5in
\epsffile[0 0 500 500]{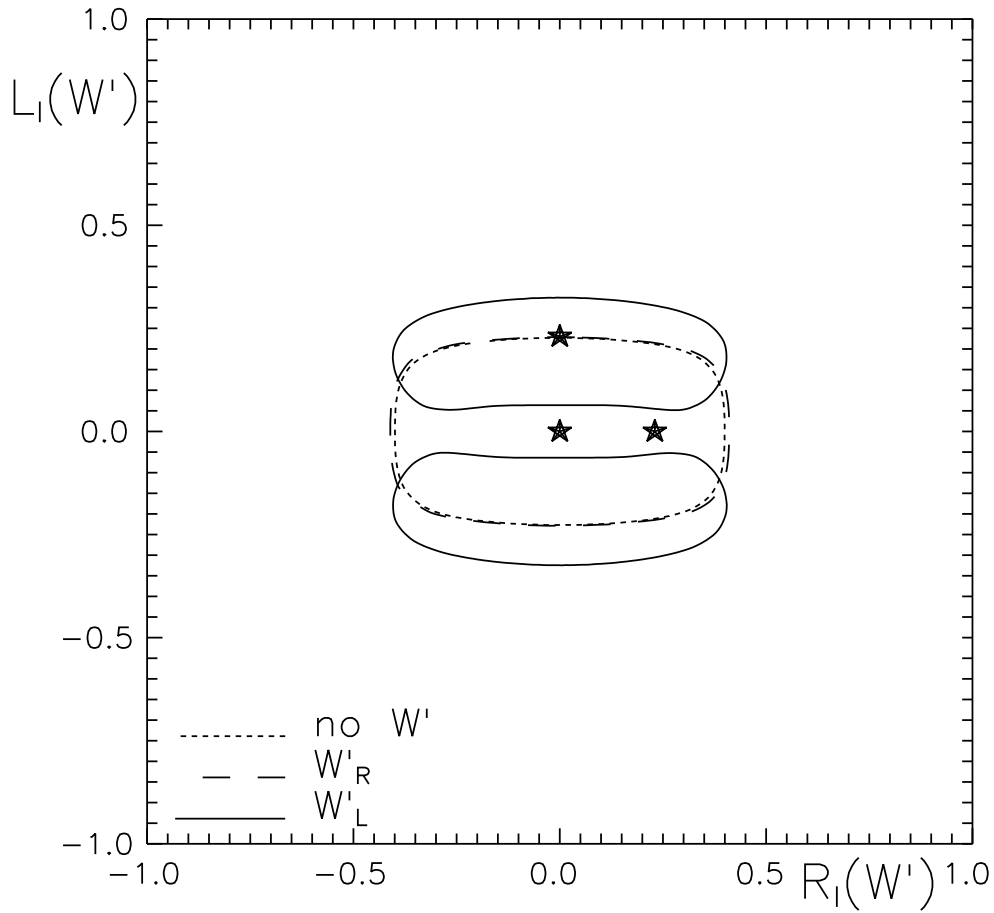}
}
\vspace{0.5in}
\caption{Constraints on the $W'$ couplings using $\sigma$ and $A_{LR}$ combined
for different $W'$ scenarios. We take
90\% electron and 60\% positron polarization,
$\sqrt{s}=0.5$ TeV, $L_{\rm int}=500$ fb$^{-1}$ and $M_{W'}=0.75$ TeV. 
A systematic error of 2\% (1\%) is included for $\sigma$ ($A_{LR}$).
The assumed models are indicated by stars.
}
\label{nnng6}
\end{figure}

\newpage
\begin{figure}
\vspace{-0.5in}
\mbox{
\epsfysize=5in
\epsffile[0 0 500 500]{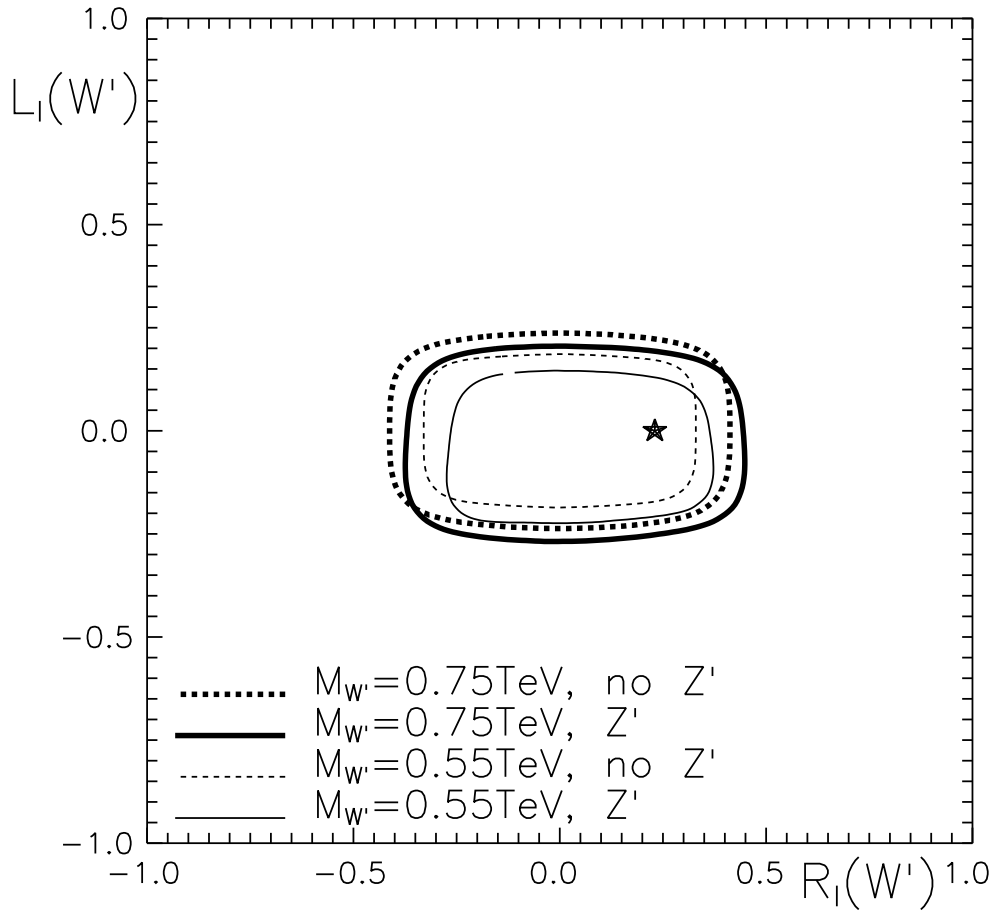}
}
\vspace{0.5in}
\caption{Constraints on the $W'$ couplings using $\sigma$ and $A_{LR}$ combined
in the LRM with $\rho=1$ and $\kappa=1$
for different $W'$ masses and different fitting strategies; see text.
We take 90\% electron and 60\% positron polarization,
$\sqrt{s}=0.5$ TeV and $L_{\rm int}=500$ fb$^{-1}$.
A systematic error of 2\% (1\%) is included for $\sigma$ ($A_{LR}$).
The assumed model (LRM) is indicated by a star.
}
\label{nnng7}
\end{figure}
\newpage
\begin{figure}
\vspace{-0.5in}
\mbox{
\epsfysize=5in
\epsffile[0 0 500 500]{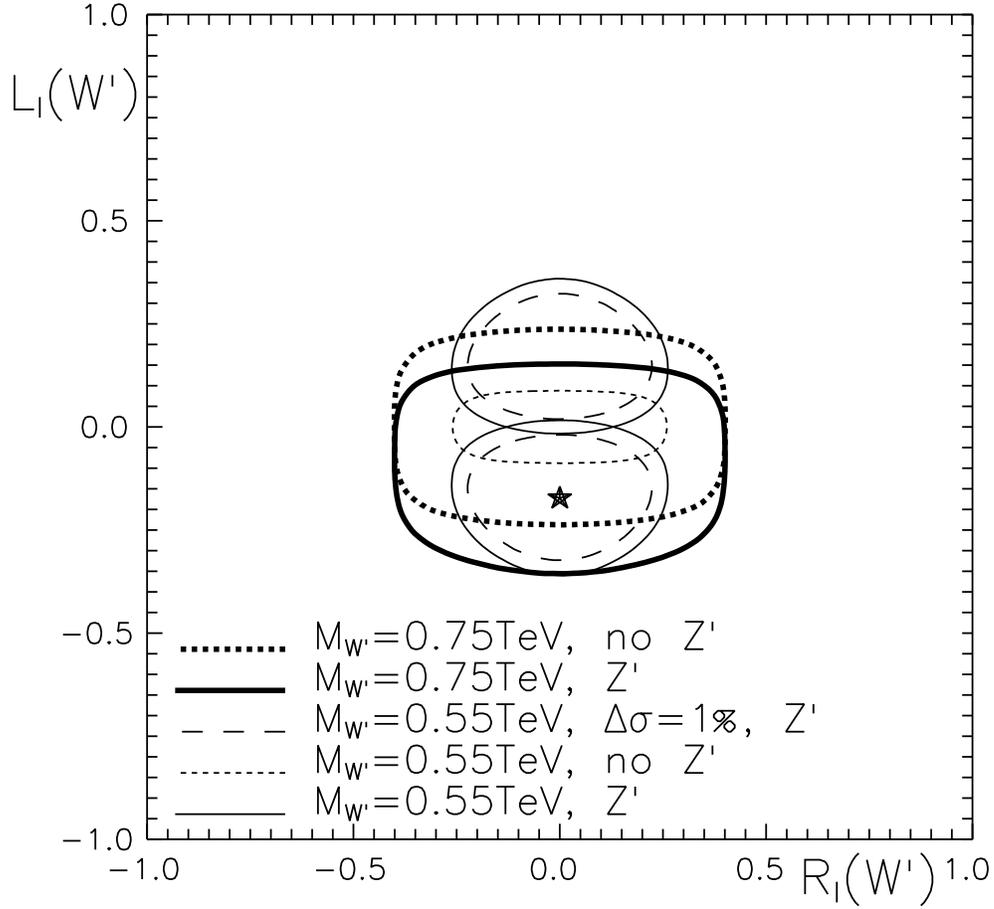}
}
\vspace{0.5in}
\caption{Constraints on the $W'$ couplings using $\sigma$ and $A_{LR}$ combined
in the UUM with $\sin\phi=0.6$
for different $W'$ masses and different fitting strategies; see text.
We take 90\% electron and 60\% positron polarization,
$\sqrt{s}=0.5$ TeV and $L_{\rm int}=500$ fb$^{-1}$.
Unless otherwise indicated,
a systematic error of 2\% (1\%) is included for 
$\sigma$ ($A_{LR}$).
The coupling of the assumed model (UUM) is indicated by a star.
}
\label{nnng8}
\end{figure}

\newpage
\begin{table}[t]
\caption{$W'$ 95\% C.L.\ discovery 
limits obtained in the SSM ($W'$), SSM ($W'+Z'$), LRM
($\kappa=\rho=1$),
UUM ($\sin\phi=0.6$), and the KK model using 
$d\sigma/dE_\gamma$ as the observable. Results are presented 
for $\sqrt{s}=500$, 1000,  and 1500 GeV and for various luminosity and
polarization scenarios, with and without a 2\% systematic error included.
For the LRM, the polarized scenario corresponds to a right-handed 
$e^-$ beam, while for all other models the beam is left-handed. 
}
\label{limitstab}
\vspace{0.4cm}
\begin{center}
\begin{tabular}{llllllllll}
           &Lum.\ (fb$^{-1}$):& 50   & 500  & 25        & 250  
                            & 50   & 500  & 25        & 250  \\
$\sqrt{s}$ &Sys. Err.:      & 0\%  & 0\%  & 0\%       & 0\% 
                            & 2\%  & 2\%  & 2\%       & 2\%  \\
(GeV)      &Model/\% pol:   & unp. & unp. & 90\%      & 90\% 
                            & unp. & unp. & 90\%      & 90\% \\
\hline
500  & SSM($W'$)    &2.4  &4.3  &2.4  &4.3  &1.55 &1.7  &1.55 &1.7  \\
     & SSM($W'+Z'$) &1.75 &3.25 &1.8  &3.25 &1.1  &1.2  &1.15 &1.25 \\
     & LRM          &0.75 &1.15 &0.85 &1.25 &0.6  &0.6  &0.75 &1.0  \\
     & UUM          &0.65 &2.1  &0.65 &2.05 &0.6  &0.6  &0.6  &0.6  \\
     & KK           &2.55 &4.55 &2.6  &4.65 &1.6  &1.75 &1.7  &1.85 \\
\hline
1000       &Lum.\ (fb$^{-1}$):& 200   & 500  & 100      & 250  
                            & 200   & 500  & 100      & 250  \\
\hline
     & SSM($W'$)    &4.2  &5.3  &4.2  &5.25 &2.15 &2.2  &2.1  &2.2  \\
     & SSM($W'+Z'$) &3.15 &4.0  &3.2  &4.1  &1.1  &1.1  &1.15 &1.45 \\
     & LRM          &1.35 &1.55 &1.35 &1.6  &0.95 &0.95 &1.25 &1.35 \\
     & UUM          &1.25 &2.45 &1.25 &2.35 &1.1  &1.1  &1.1  &1.1  \\
     & KK           &4.55 &5.75 &4.6  &5.85 &2.15 &2.2  &2.25 &2.3  \\
\hline
1500       &Lum.\ (fb$^{-1}$):& 200   & 500  & 100      & 250  
                            & 200   & 500  & 100      & 250  \\
\hline
     & SSM($W'$)    &4.7  &5.95 &4.65 &5.85 &2.45 &2.55 &2.45 &2.55 \\
     & SSM($W'+Z'$) &3.4  &4.45 &3.45 &4.5  &1.45 &1.45 &1.55 &1.55 \\
     & LRM          &1.65 &1.9  &1.7  &1.9  &1.3  &1.3  &1.55 &1.65 \\
     & UUM          &1.8  &1.85 &1.8  &1.85 &1.55 &1.55 &1.55 &1.55 \\
     & KK           &5.05 &6.45 &5.1  &6.45 &2.35 &2.45 &2.45 &2.55 \\
\end{tabular}
\end{center}
\end{table}

\end{document}